\documentclass[preprintnumbers,superscriptaddress,nofootinbib]{revtex4}
\usepackage{graphicx}
\def\hsm{h_{SM}}
\def\mhsm{m_{\hsm}}
\def\mw{m_W}

\def\wp{W^+}
\def\wm{W^-}
\def\anti{\overline}
\def\gev{~{\rm GeV}}
\def\tev{~{\rm TeV}}

\def\lphi{\Lambda_\phi}

\def\mphi{m_\phi}

\def\hbar{\overline h}
\def\lam{\lambda}

\def\h{h}
\def\mh{m_{\h}}
\def\fbi{~{\rm fb}^{-1}}
\def\tanb{\tan\beta}
\def\lam{\lambda}

\def\hh{H}
\def\hl{h}
\def\mhh{m_{\hh}}
\def\mhl{m_{\hl}}
\def\ha{A}
\def\mha{m_{\ha}}
\def\gam{\gamma}
\def\epem{e^+e^-}
\def\gfvh{g_{fV\h}}
\def\gfvphi{g_{fV\phi}}
\def\gama{\gam a}
\def\gamb{\gam b}
\def\Fig#1{Fig.~\ref{#1}}

\def\anti{\overline}
\def\gam{\gamma}

\def\ifmath#1{\relax\ifmmode #1\else $#1$\fi}

\def\lsim{\mathrel{\raise.3ex\hbox{$<$\kern-.75em\lower1ex\hbox{$\sim$}}}}
\def\gsim{\mathrel{\raise.3ex\hbox{$>$\kern-.75em\lower1ex\hbox{$\sim$}}}}

    \def\fillboxx#1#2{\hbox to #1{\vbox to #2{\vfil}\hfil}    }

\def\vev#1{\langle #1 \rangle}

\def\tanb{\tan\beta}

\def\mw{m_W}

\def\wp{W^+}
\def\wm{W^-}

\def\h{h}
\def\mh{m_{\h}}

\def\lam{\lambda}

\def\gam{\gamma}

\def\anti{\overline}
\def\epem{e^+e^-}

\def\rtsee{\sqrt s_{ee}}

\def\anti{\overline}
\def\wp{W^+}
\def\wm{W^-}
\def\mw{m_W}

\def\h{h}
\def\mh{m_{\h}}

\def\hsm{h_{SM}}
\def\mhsm{m_{\hsm}}

\def\hl{h^0}
\def\mhl{m_{\hl}}

\def\ha{A^0}
\def\mha{m_{\ha}}

\def\hh{H^0}
\def\mhh{m_{\hh}}

\def\fbi{~{\rm fb}^{-1}}

\def\gev{~{\rm GeV}}
\def\tev{~{\rm TeV}}

\def\MPL #1 #2 #3 {{ Mod.~Phys.~Lett.}~{\bf#1} (#3) #2}
\def\NPB #1 #2 #3 {{ Nucl.~Phys.}~{\bf #1} (#3) #2}
\def\PLB #1 #2 #3 {{ Phys.~Lett.}~{\bf #1} (#3) #2}
\def\PR #1 #2 #3 {{ Phys.~Rep.}~{\bf#1} (#3) #2}
\def\PRD #1 #2 #3 {{ Phys.~Rev.}~{\bf #1} (#3) #2}
\def\PRL #1 #2 #3 {{ Phys.~Rev.~Lett.}~{\bf#1} (#3) #2}
\def\RMP #1 #2 #3 {{ Rev.~Mod.~Phys.}~{\bf#1} (#3) #2}
\def\ZPC #1 #2 #3 {{ Z.~Phys.}~{\bf #1} (#3) #2}
\def\IJMP #1 #2 #3 {{ Int.~J.~Mod.~Phys.}~{\bf#1} (#3) #2}
\def\NIM #1 #2 #3 {{ Nucl.~Inst.~and~Meth.}~{\bf#1} {#3} #2}
\def\JHEP #1 #2 #3 {{ JHEP}~{\bf#1} (#3) #2}
\newcommand{\nc}{\newcommand}
\nc{\beq}{\begin{equation}}   \nc{\eeq}{\end{equation}}
\nc{\bea}{\begin{eqnarray}}   \nc{\eea}{\end{eqnarray}}
\nc{\baa}{\begin{array}}      \nc{\eaa}{\end{array}}
\nc{\bit}{\begin{itemize}}    \nc{\eit}{\end{itemize}}
\nc{\bed}{\begin{description}}    \nc{\eed}{\end{description}}
\nc{\ben}{\begin{enumerate}}  \nc{\een}{\end{enumerate}}
\nc{\bce}{\begin{center}}     \nc{\ece}{\end{center}}
\def\beqa{\begin{eqnarray}}
\def\eeqa{\end{eqnarray}}
\def\lsim{\mathrel{\raise.3ex\hbox{$<$\kern-.75em\lower1ex\hbox{$\sim$}}}}
\def\gsim{\mathrel{\raise.3ex\hbox{$>$\kern-.75em\lower1ex\hbox{$\sim$}}}}
\def\vev#1{\langle #1 \rangle}
\def\lam{\lambda}

\newcommand{\gghh}{\ensuremath{\gamma \gamma \rightarrow H^+ H^-}}
\newcommand{\GeV}{\ensuremath{\mathrm{GeV}}}

\newcommand{\hpm}{\ensuremath{H^{\pm}}}
\newcommand{\mhpm}{\ensuremath{m_{H^{\pm}}}}
\newcommand{\GeVcsq}{\ensuremath{\mathrm{GeV/c^2}}}

\newcommand{\ggtt}{\ensuremath{\gamma \gamma \rightarrow \tau^+ \tau^-}}

\newcommand{\fb}{\ensuremath{\mathrm{fb}}}
\newcommand{\pb}{\ensuremath{\mathrm{pb}}}
\newcommand{\ggww}{\ensuremath{\gamma \gamma \rightarrow W^+ W^-}}
\newcommand{\ggwwtt}{\ensuremath{\gamma \gamma \rightarrow W^+ W^- \rightarrow \tau^+\tau^-}}
\newcommand{\htt}{\ensuremath{H^+ \rightarrow \tau^+ \nu_{\tau}}}
\def\lesssim{\raisebox{-.5ex}{\,\rlap{$\sim$}} \raisebox{.4ex}{$<$}\;} 
\def\moresim{\raisebox{-.5ex}{\,\rlap{$\sim$}} \raisebox{.4ex}{$>$}\;} 
\newcommand{\gaga}                 {\gamma\gamma}
\newcommand{\pwdgaga}              {\Gamma_{\gamma\gamma}}
\newcommand{\bbbar}                {b\bar{b}}
\newcommand{\epsega}               {\epsilon_{e\rightarrow\gamma}}
\newcommand{\Mgaga}                {M_{\gamma\gamma}}
\begin{document}

%\preprint{FERMILAB-Conf-02/nnn-T}
%\preprint{hep-ph/0208219}
\preprint{nuhep-exp/02-012\cr UCD-02-11}

\renewcommand{\thefootnote}{\fnsymbol{footnote}}

\title{New results for a photon-photon collider}

\author{David Asner}
\affiliation{Lawrence Livermore National Laboratory, Livermore, CA}

\author{Bohdan Grzadkowski}
\affiliation{Institute of Theoretical Physics, Warsaw University, Warsaw, Poland}

\author{John F. Gunion}
\affiliation{University of California, Davis, California 95616, USA}

\author{Heather~E.~Logan}
\affiliation{Theoretical Physics Department, Fermilab, PO Box 500, Batavia, Illinois 60510-0500, USA}

\author{Victoria Martin}
\affiliation{Northwestern University, Evanston, Illinois, USA}

\author{Michael Schmitt}
\affiliation{Northwestern University, Evanston, Illinois, USA}

\author{Mayda M. Velasco}
\affiliation{Northwestern University, Evanston, Illinois, USA}

\date{\today}

\begin{abstract}
We present new results from studies in progress on physics at a
two-photon collider.  We report on the sensitivity to top squark 
parameters of MSSM Higgs boson production in two-photon collisions;
Higgs boson decay to two photons;
radion production in models of warped extra dimensions; chargino pair 
production; sensitivity to the trilinear Higgs boson coupling;
charged Higgs boson pair production; and we discuss the backgrounds
produced by resolved photon-photon interactions.
\end{abstract}

\maketitle

%%%%%%%%%%%%%%%%%%%%%%%%%%%%%%%%%%%%%%%%%%%%%%%%%%%%%%%%%%%%%%%%%%%%%%%%

\renewcommand{\thefootnote}{\arabic{footnote}}
\setcounter{footnote}{0}

%%%%%%%%%%%%%%%%%%%%%%%%%%%%%%%%%%%%%%%%%%%%%%%%%%%%%%%%%%%%%%%%%%%%%%%%%%%%
\pagestyle{plain}

\section{Introduction}
\par
This document summarizes recent results obtained by the American
working group on $\gaga$~colliders, and is intended to be a supporting
document for discussions at {\sc LC2002}, to be held on Jeju Island, South~Korea.
\par
We have been considering two machine scenarios.  In the first, a `Higgs-factory'
is based on high energy photon beams optimized for the $s$-channel production
of a Standard Model-like Higgs boson of mass in the range 100--130~GeV.
The properties of this Higgs boson would be explored in a dedicated fashion.
Paramount among these is the measurement of $\pwdgaga$, which is sensitive
to new charged particles such as the top squark, and also the top Yukawa coupling.  
Certain rare decays are available due to low backgrounds, such as $h\rightarrow\gaga$,
$h\rightarrow\gamma Z$, and $h\rightarrow Z\,Z^*$.  For an earlier study
and description of the machine, see Ref.~\cite{AsnerCliche}.
In the second, more conventional scenario, a $\gaga$~collider runs in
parallel with a high energy $\epem$ linear collider.  In this case the
study of the heavier neutral Higgs bosons, $H$ and $A$, can be pursued
even in cases in which they are not visible in the companion $\epem$ machine~\cite{AsnerNLC}.
Linearly polarized beams provide information about the charge-parity~(CP)
nature
of these particles, so in this and in other senses a $\gaga$~collider
complements the $\epem$ machine.  As discussed in this document, effective
cross sections are as large as or larger than the corresponding $\epem$
cross sections.  For example, we have studied the copious and clean signal
from $\gaga\rightarrow H^+H^-$.  Also, it appears that the sensitivity
of a high energy $\gaga$~collider to the Higgs self-coupling is about the
same as the $\epem$ machine.  A similar conclusion is reached for chargino
production.  While one would not advocate substituting a high energy
$\epem$ collider by a $\gaga$~one, it is clear that more will be understood
about the Higgs (and other) sector if both machines are in operation.
\par
This paper is organized as follows.
In Sec.~\ref{sec:hstop} we describe how the mass of the heavier top 
squark can be inferred from a measurement of the $h^0 \gamma \gamma$ 
coupling in the minimal supersymmetric Standard Model (MSSM), if the lighter top squark mass and mixing angle 
are known. 
In Sec.~\ref{sec:hgg} we examine Higgs boson decay to two photons.
In Sec.~\ref{sec:radion} we examine radion production in two-photon
collisions in models of warped extra dimensions.
In Sec.~\ref{sec:charginos} we study chargino pair production in
two-photon collisions and estimate the event yield.
In Sec.~\ref{sec:trilinear} we study pair production of the Standard Model~(SM)
Higgs
boson in two-photon collisions, which is sensitive to the trilinear
Higgs boson coupling.
In Sec.~\ref{sec:H+H-} we study signal and background to charged Higgs
boson pair production in two-photon collisions, with decays to taus.
In Sec.~\ref{sec:resolved} we examine the background at a two-photon
collider due to resolved photons.
Finally we summarize our results in Sec.~\ref{sec:summary}.

%----------------------------------------------------------------
\section{\label{sec:hstop}Top squark parameters from MSSM Higgs 
boson production in photon-photon collisions}

%{\it Talk presented by ...}

A two-photon collider optimized to run on a light Higgs boson ($h^0$) 
$s$-channel resonance can measure the rate of 
$\gamma\gamma \to h^0 \to b \bar b$ with a precision of about 2\%
\cite{Maydagg,AsnerCliche,AsnerNLC,Soldner}.  
Combining this with the expected model-independent measurement of 
BR($h^0 \to b \bar b$) of 2-3\% from an $e^+e^-$ collider
\cite{Battaglia,eeBRb}, one can extract the $h^0\gamma\gamma$ coupling
with a precision of 2-3\%.
In the MSSM, such a measurement is sensitive to contributions 
to the loop-induced $h^0\gamma\gamma$ coupling from supersymmetric~(SUSY)
particles, mainly top squarks and charginos \cite{Djouadigg}, 
and to deviations in
the couplings of $h^0$ to SM particles from their SM values.
We show that this measurement of the 
$h^0\gamma\gamma$ coupling can be combined with $e^+e^-$ collider
data on the lighter top squark mass and mixing angle
to constrain the mass of the heavier top squark to within about
$\pm 20$ GeV or less, if the lighter top squark is not too heavy.

In the MSSM, there can generally be a large splitting between
the masses of the lighter ($\tilde t_1$) and heavier ($\tilde t_2$) 
top squarks, and $\tilde t_1$ can be quite light.
Thus, one can envision a situation in which $\tilde t_1$ can
be pair-produced at a first-stage (500 GeV) $e^+e^-$ collider, but
$\tilde t_2$ is kinematically inaccessible.  In this case,
$m_{\tilde t_1}$ can be measured to high precision.  
The top squark mixing angle $\cos^2\theta_{\tilde t}$ can 
also be measured at the per cent level because of the dependence
of the $e^+e^- \to \tilde t_1 \tilde t_1^*$ cross
section on this angle.

In such a situation one would like to constrain the mass of $\tilde t_2$ 
indirectly.  In the decoupling limit of large $m_A$,
the $h^0 \tilde t_i \tilde t_j^*$ couplings can be written solely
in terms of known SM parameters and 
$m_{\tilde t_1}$, $m_{\tilde t_2}$ and $\cos \theta_{\tilde t}$,
with a very mild additional dependence on $\cos 2\beta$. 
A measurement of one of these couplings then allows one to 
extract $m_{\tilde t_2}$.
The cross sections for
$e^+e^- \to \tilde t_1 \tilde t_1^* h^0$ \cite{Belangertth} and
$e^+e^- \to \tilde t_1 \tilde t_1^* Z$ \cite{BelangerttZ} are sensitive
to the $h^0 \tilde t_1 \tilde t_1^*$ coupling and hence to $m_{\tilde t_2}$.  
This method succeeds only if $\tilde t_1$ is light enough and 
$\cos\theta_{\tilde t}$ is moderate (i.e., not close to 1 or 0).
For $m_{\tilde t_1} = 120$ GeV and $\cos\theta_{\tilde t} = 0.4$,
the sensitivity to $m_{\tilde t_2}$ is about $\pm 20$ GeV 
from $e^+e^- \to \tilde t_1 \tilde t_1^* h^0$ \cite{Belangertth}.
This method is no longer effective if the $h^0 \tilde t_1 \tilde t_1^*$
coupling is small, since the cross section becomes too small to detect.

\begin{figure}
\resizebox{\textwidth}{!}{
\rotatebox{270}{\includegraphics{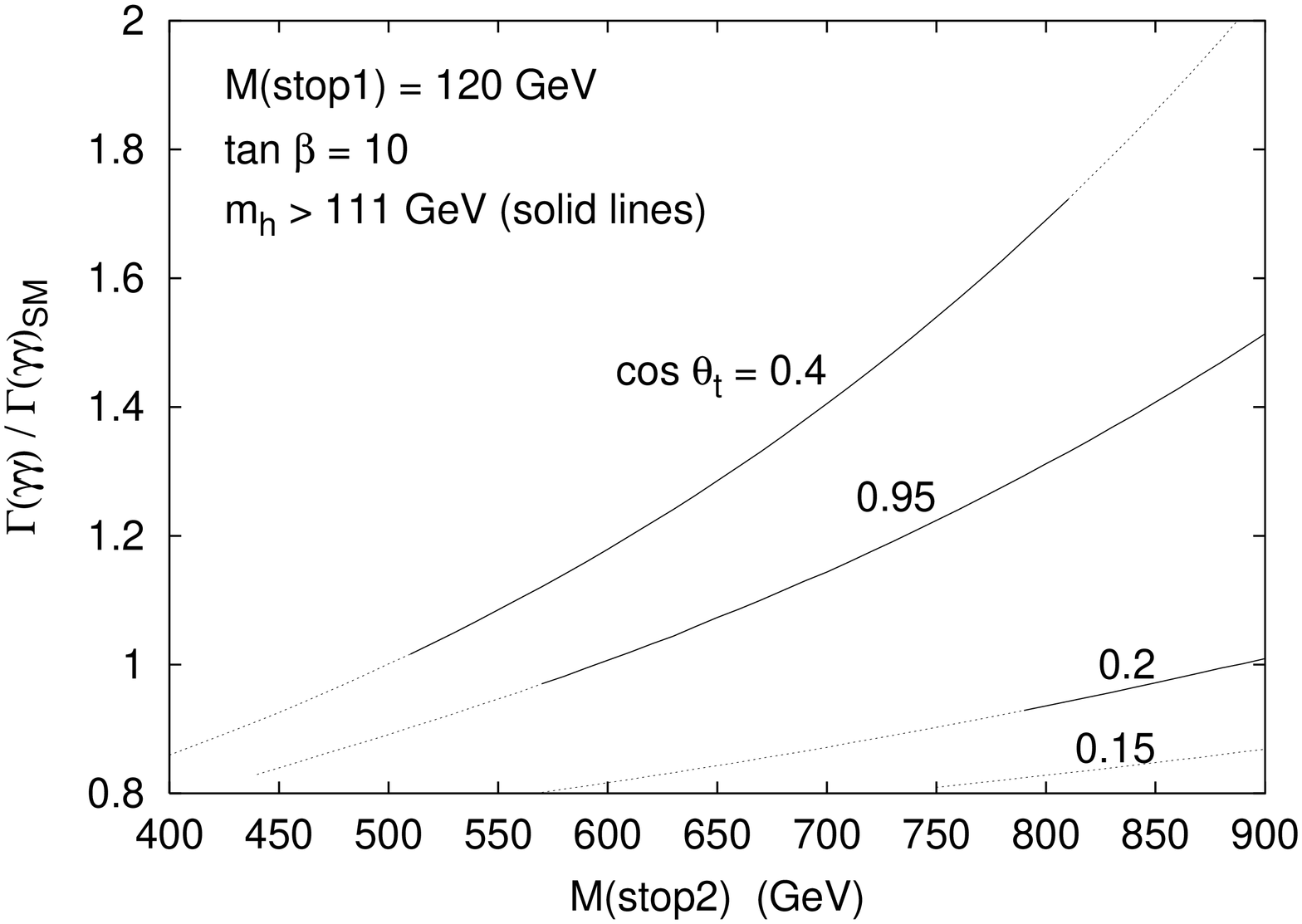}}
\rotatebox{270}{\includegraphics{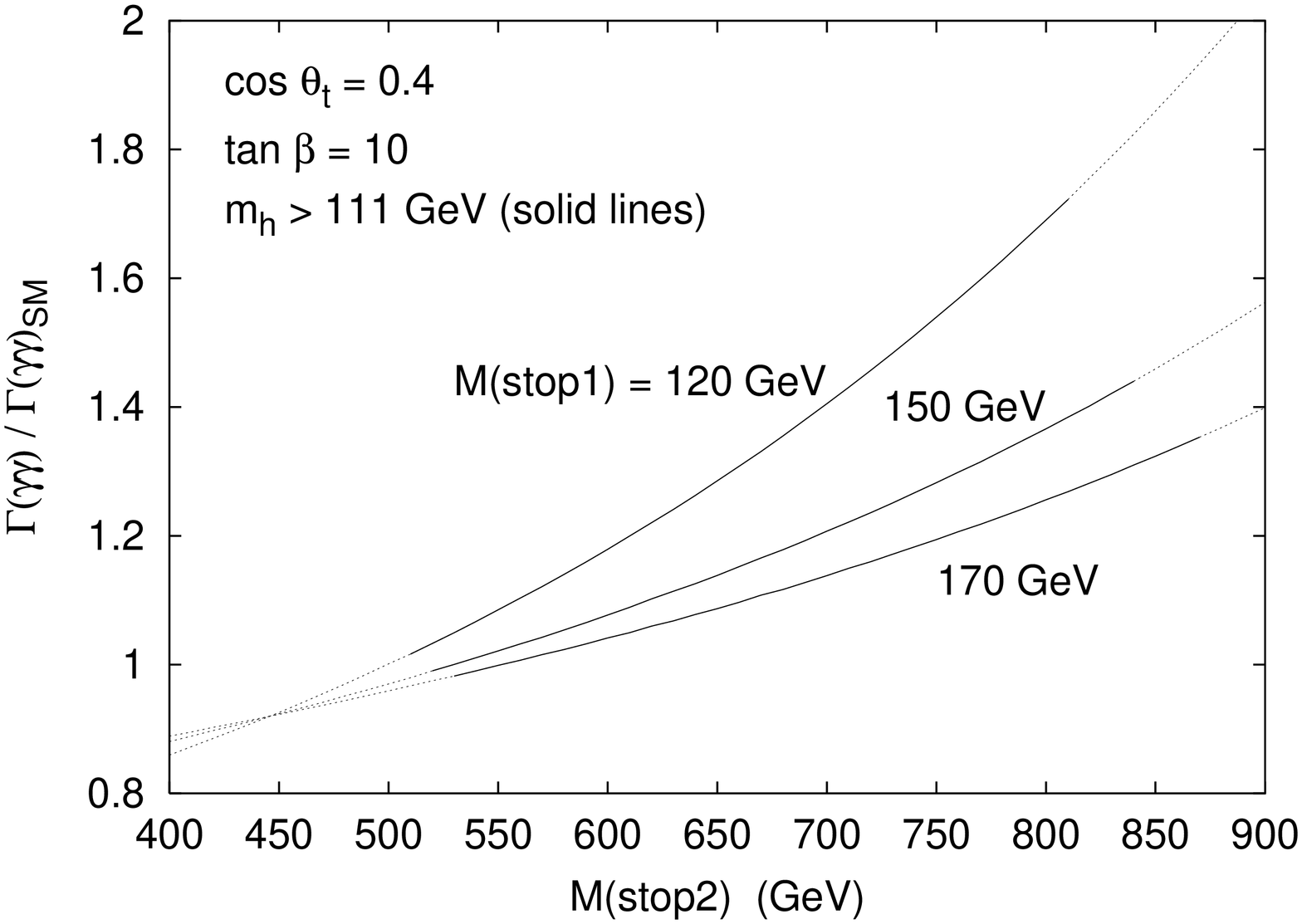}}}
\caption{Dependence of the partial width $\Gamma(h^0 \to \gamma\gamma)$ 
on $m_{\tilde t_2}$ for various values of $m_{\tilde t_1}$ and 
$\cos\theta_{\tilde t}$.  Here $m_A = 1$ TeV, $\tan\beta = 10$, 
$M_2 = -\mu = 200$ GeV,
and the remaining SUSY mass parameters are set to 1 TeV.  
The numerical calculations were done
using HDECAY \cite{HDECAY}.}
\label{fig:hggstop2}
\end{figure}

Here we study the sensitivity to $m_{\tilde t_2}$
of the $h^0 \to \gamma\gamma$ partial width,
assuming that $m_{\tilde t_1}$ and $\cos\theta_{\tilde t}$
have been measured in $e^+e^-$ collisions.  
This is shown in Fig.~\ref{fig:hggstop2} for various values of 
$m_{\tilde t_1}$ and $\cos\theta_{\tilde t}$, in the large-$m_A$ limit.
Using the knowledge of $m_{\tilde t_1}$ and $\cos\theta_{\tilde t}$ from the
$e^+e^-$ collider, a 2\% measurement of $\Gamma(h^0 \to \gamma\gamma)$
allows one to extract $m_{\tilde t_2}$ with a statistical
precision of about $\pm 10$ GeV for $m_{\tilde t_1} = 120$ GeV and 
$\cos\theta_{\tilde t} = 0.4$.
This precision worsens as $m_{\tilde t_1}$
increases or $\cos\theta_{\tilde t}$ varies, because the dependence of
$\Gamma(h^0 \to \gamma\gamma)$ on $m_{\tilde t_2}$ becomes weaker (see 
Fig.~\ref{fig:hggstop2}).
The soft SUSY-breaking mass parameters $M_{Q_3}$, $M_{U_3}$ and 
$X_t \equiv A_t - \mu \cot\beta$ can then
be inferred from $m_{\tilde t_1}$, $\cos\theta_{\tilde t}$ and $m_{\tilde t_2}$.

Additional uncertainties arise due to the dependence on $m_A$ of the 
$h^0 \tilde t_i \tilde t_j^*$ couplings and additional chargino loop 
contributions to
$\Gamma(h^0 \to \gamma\gamma)$. To a lesser extent, the 
bottom squarks also contribute at large $\tan\beta$ \cite{Djouadisb}.
In most of the MSSM parameter space, values of $m_A$ below
about 600 GeV will lead to at least a $2 \sigma$ deviation in the
ratio ${\rm BR}(h^0 \to b \bar b)/{\rm BR}(h^0 \to WW)$, measurable
in either $e^+e^-$ \cite{Battaglia,CHLM} or $\gamma\gamma$ \cite{Maydagg} 
collisions.  For larger $m_A$, the effect on
$\Gamma(h^0 \to \gamma\gamma)$ is negligible.  
Values of $m_A$ as low as 200 GeV lead to an additional $\pm 4$ GeV 
parametric uncertainty in $m_{\tilde t_2}$ for the parameters considered above.

The uncertainty due to chargino loops can be removed if the 
parameters of the chargino mass matrix are already known.
The measurements are straightforward at an $e^+e^-$ collider~\cite{GudiCharginos}.  
In fact, complete 
reconstruction of the chargino and neutralino mass matrices is
possible even if only $\tilde \chi_1^+$, $\tilde \chi_1^0$ and 
$\tilde \chi_2^0$ are kinematically accessible \cite{GudiPartial}.
If $\tilde \chi_1^+$ is too heavy to be pair produced in 500 GeV $e^+e^-$ 
collisions, its effect on $\Gamma(h^0 \to \gamma\gamma)$ is also diminished
due to decoupling.  For $M_2 = -\mu = 250$ GeV (leading to a 
$\tilde \chi_1^+$ mass of 200-214 GeV, depending on $\tan\beta$), the unknown
chargino contribution leads to an additional $\pm 8$ GeV uncertainty
in $m_{\tilde t_2}$ for the parameters considered above, assuming no
knowledge of the value of $\tan\beta$.  If $\tan\beta$ is large, the 
chargino contribution is quite small and leads to far less uncertainty.

Finally, we comment briefly on Higgs boson decays.
Decays of $h^0$ into $b \bar b$, $WW$, $ZZ$, $\gamma\gamma$ and $Z\gamma$
can be measured at a two-photon collider.
The theoretical behavior of these decay widths in the MSSM has been analyzed
thoroughly \cite{CHLM,Djouadigg,DjouadiZg}.
Neglecting contributions of SUSY particles in the loop, 
the ratio $\Gamma(h^0 \to Z \gamma)/\Gamma(h^0 \to WW)$ approaches its
SM value very quickly with increasing $m_A$; the deviation from the SM
prediction is less than 10\% for $m_A > 130$ GeV and less than 2\% for
$m_A > 150$ GeV.  Thus a measurement of this ratio can be used as a test 
for light charged non-SM particles in the $h^0 \to Z \gamma$ loop.
Ref.~\cite{DjouadiZg} found that top squarks can contribute up to 5\% 
deviations for $m_{\tilde t_1} > 100$ GeV and charginos can contribute
up to 10\% deviations for $m_{\tilde \chi_1^+} > 100$ GeV.

%-----------------------------------------------------------
\section{\label{sec:hgg}Higgs Boson decay to two photons}
\par
The partial width for $h\rightarrow\gaga$, $\pwdgaga$, is sensitive to
charged non-SM particles.  For example, in the MSSM with a light and a
heavy stop, the mass of the heavier stop can be inferred from the
measured value of $\pwdgaga$ if the mass of the lighter stop and the
mixing angle are already known -- see Section~\ref{sec:hstop}.
Since the Higgs cross section is proportional to $\pwdgaga$,
the rate $\gaga\rightarrow\bbbar$ when combined with a
measurement of $BR(h\rightarrow\bbbar)$ from an $\epem$ collider
leads to a 2--3\% measurement of $\pwdgaga$~\cite{AsnerCliche,Maydagg}.
\par
The channel $h\rightarrow\gaga$ is doubly sensitive to $\pwdgaga$.
In an $\epem$ experiment, this channel can be difficult to observe
due to large backgrounds.  But in a $\gaga$~collider, the backgrounds
are low because continuum scattering $\gaga\rightarrow\gaga$ proceeds
only through loops.  Preliminary studies have shown that a clear
signal can be extracted with a very simple analysis, provided the
calorimeter performance is good enough.  Furthermore, a direct
measurement of the Higgs mass is possible from the $\gaga$ invariant
mass distribution.  We have continued these studies and made them
more realistic, as briefly described here; the conclusions remain
positive.
\par
\begin{figure}
\resizebox{.5\textwidth}{!}{
\rotatebox{0}{\includegraphics{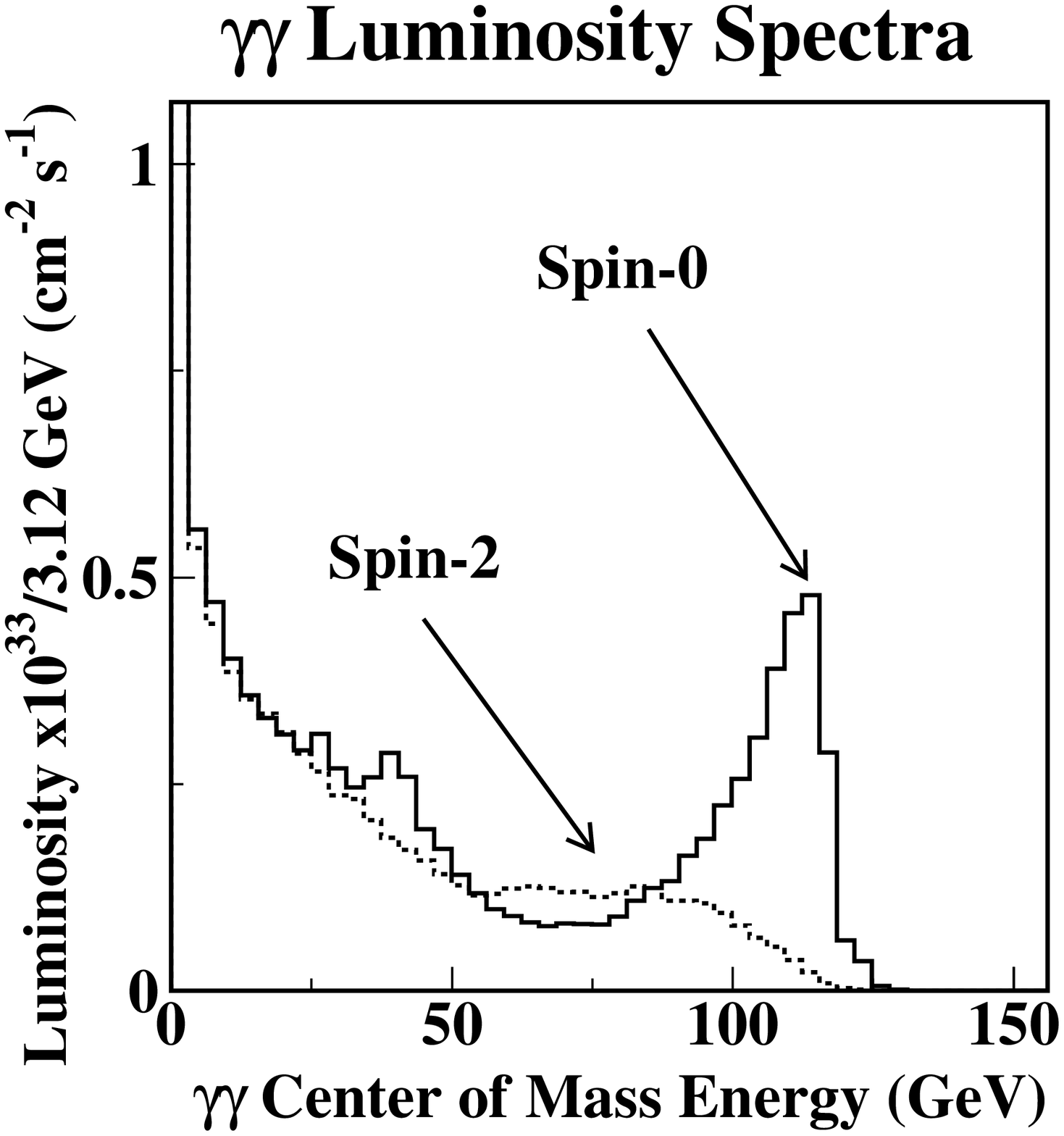}}
\rotatebox{0}{\includegraphics{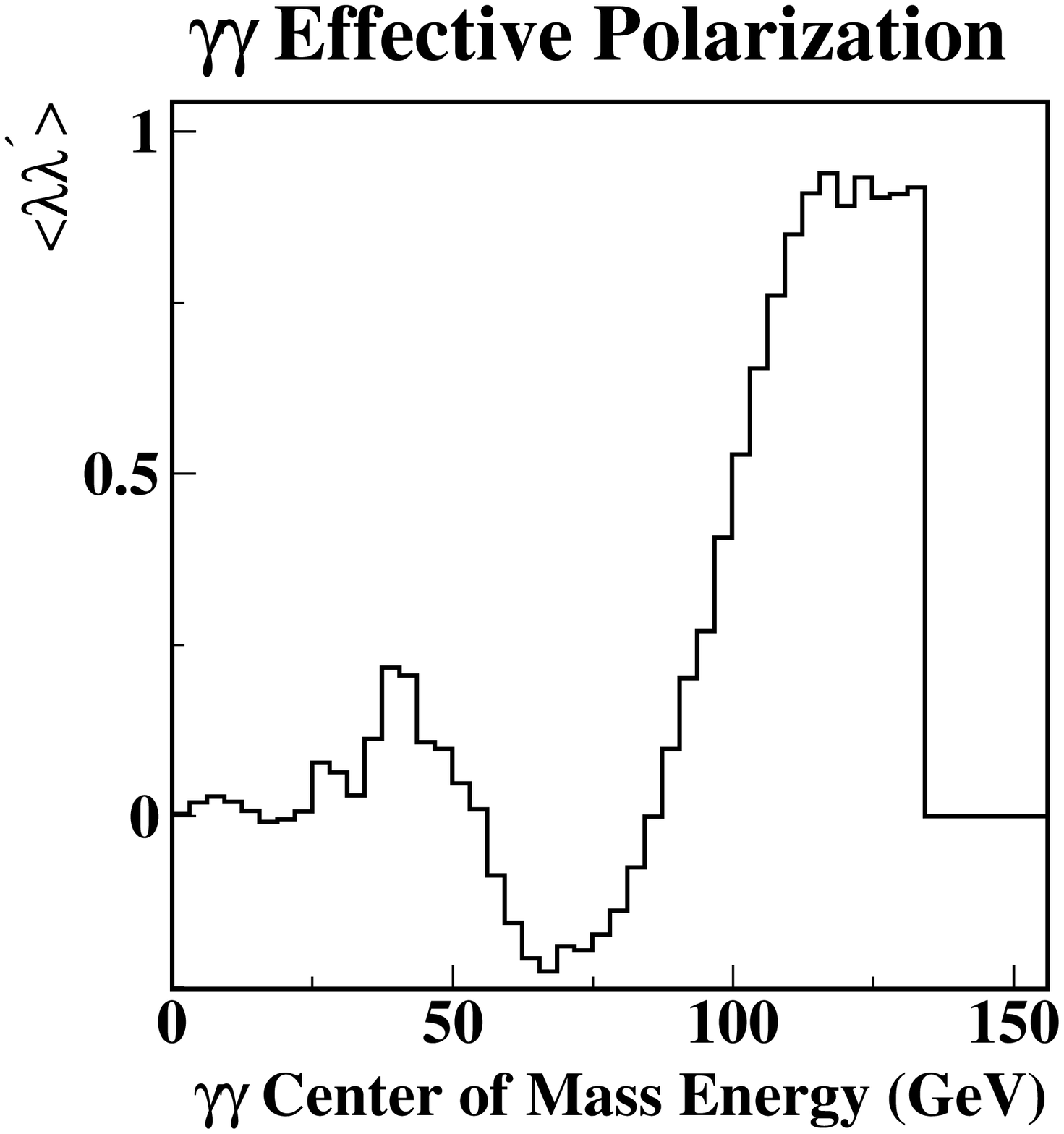}}}
\caption[0]{Luminosity for a $10^7$~sec year and
associated expectation value for the product of photon polarizations, $\vev{\lam\lam'}$, are plotted for $\rtsee=150$~GeV
($x=4.1$ for $1.054/3~\mu$m laser wavelength),
assuming 80\% electron beam polarizations.}
\label{fig:cliclums}
\end{figure}

This study assumes a Higgs-factory machine, running with the
peak of the $\gaga$~energy spectrum at the Higgs mass. For $m_h = 120$~GeV,
this corresponds to 75~GeV electron beams~\cite{AsnerCliche}.
The beam parameters for this facility are given in Table~\ref{tab:beampar}
and the corresponding luminosity and polarization spectra in 
Fig.~\ref{fig:cliclums}.
The assumed integrated luminosity is sufficient to produce 
11,400~Higgs bosons.  The SM branching ratio for $m_h = 120$~GeV
is $2.1\times 10^{-3}$.
\begin{table}[h]
\begin{center}
\begin{tabular}[c]{ccc}
\hline
\hline
Electron Beam Energy (GeV) & 75 & 250  \\
$\beta_x/\beta_y$~(mm) & 2/0.020 & 4/0.065   \\
$\epsilon_x/\epsilon_y~(\times 10^{-8})$ & 140/5 & 360/7.1 \\
$\sigma_x/\sigma_y ~({\rm nm})$ & 138/2.6 & 172/3.1  \\
$\sigma_z$~(microns) & 30 &  156  \\
$ N~(\times 10^{10})$ & 0.4 & 1.5  \\
$ e^-~{\rm Polarization}~ (\%)$ & 80 & 80 \\
repetition rate (Hz) & $100\times 154 \times 11$ & 120$\times$95  \\
Laser Pulse Energy (J) & 1.0/70\%=1.4 & 1.0 \\
Laser $\lambda$ (microns) & 1.054/3 = 0.351 &  1.054  \\
CP-IP distance (mm) & 1 & 2  \\
\hline
\hline
\end{tabular}
\caption{Laser and electron beam parameters for ${\sqrt s_{ee}} = 150$~GeV
and ${\sqrt s_{ee}} = 500$~GeV. The beam parameters for ${\sqrt s_{ee}} = 500$~GeV
differ from the NLC-$e^+e^-$ parameters. The bunch charge has been
doubled to improve luminosity. Consequently both the vertical emittance, 
$\epsilon_y$, and the bunch length, $\sigma_z$ are increased. Additionally,
the total current is conserved as the repetition rate reduced by a factor of 2.
The optimal laser wavelength decreases as the beam energy decreases.
We assume that non-linear optics are used to triple the laser frequency for
the ${\sqrt s_{ee}} = 150$~GeV machine and that this procedure is 70\% efficient,
thus more laser power is required.}
\label{tab:beampar}
\end{center}
\end{table}
\par
The first improvement is the introduction of a calorimeter geometry
and simulation.  An aggressive design has been assumed, with coverage
down to $3^\circ$ and a high degree of granularity.  The energy 
resolution is the same as the CMS design, namely, 
$\sigma_E/E = \left( (0.015/\sqrt{E})^2 + (0.0045)^2 \right)^{1/2}$.
A cone algorithm is used to find the high energy photon showers.
Threshold and clustering effects have little impact on the
mass resolution, but do lead to a shift in the observed mass
of~0.25~GeV.  The width of the cone, $\Delta R = 0.3$, was chosen
as small as possible without impacting the signal resolution.
\par
\begin{figure}[t]
\begin{center}
%\mbox{
\resizebox{0.75\textwidth}{!}{
\includegraphics{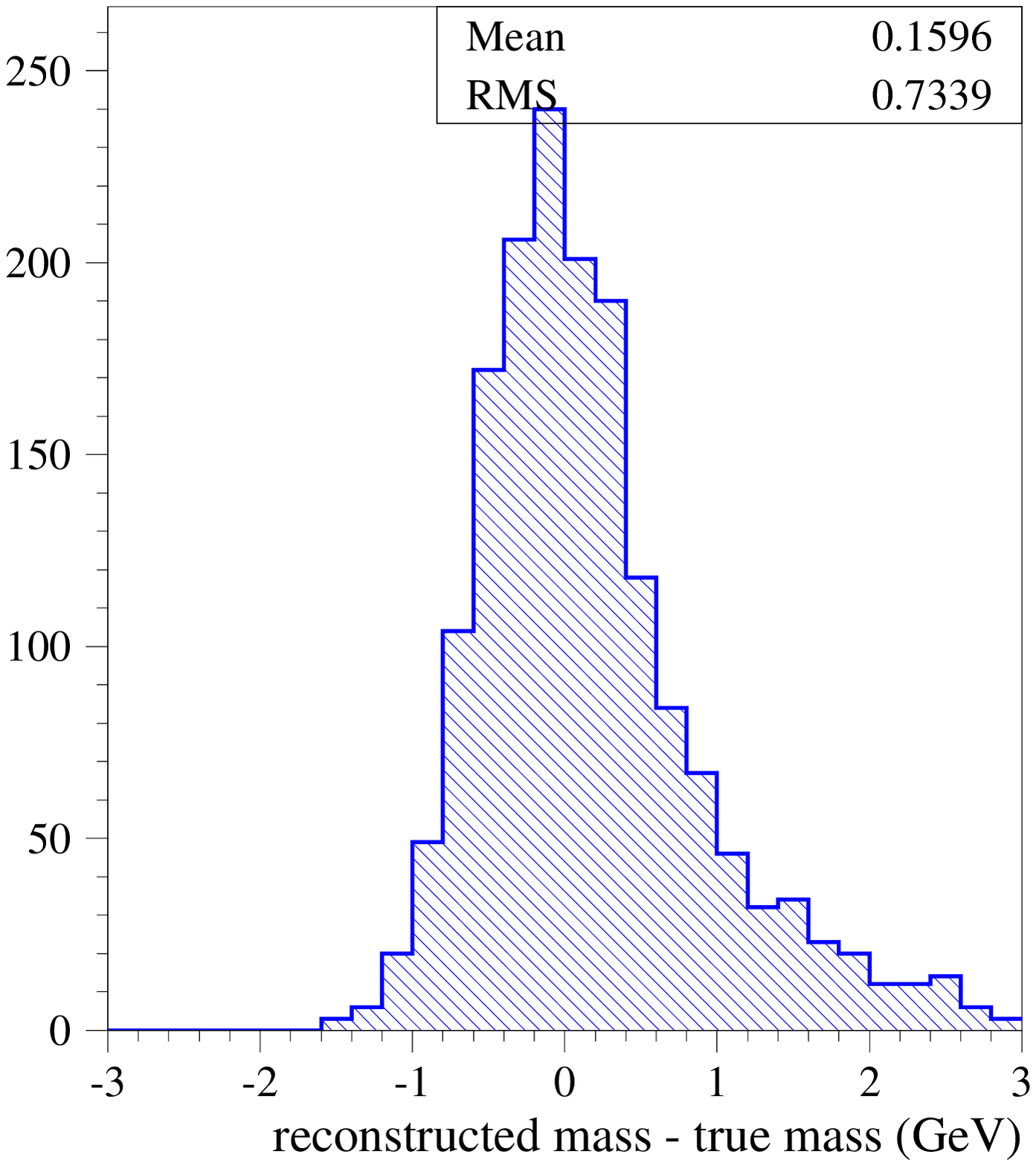}\hspace*{0.25in}
\includegraphics{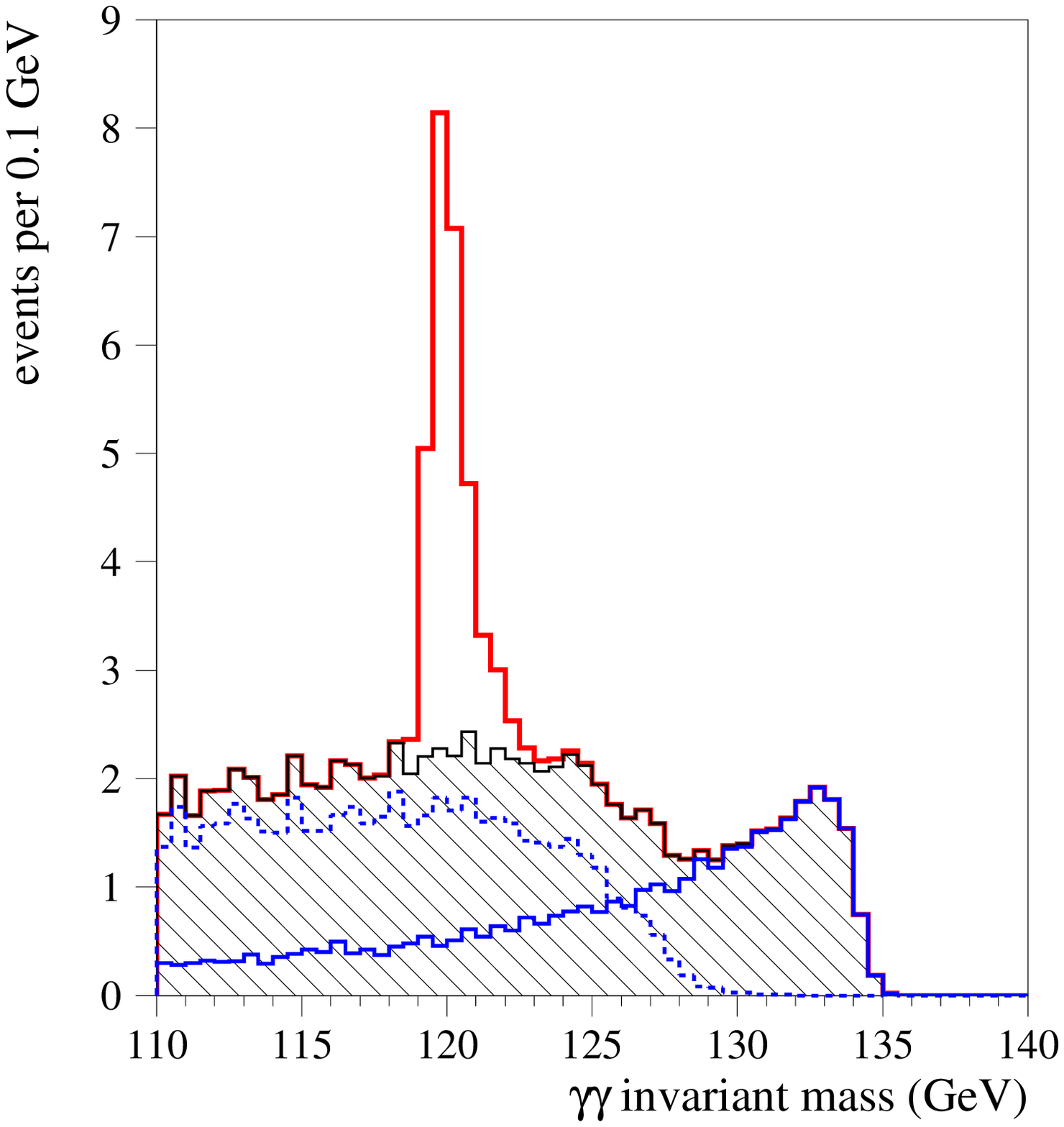}
%}
}
\caption{\label{fig:hggrecmass}
(left) The reconstructed Higgs mass compared to the true mass.
(right) The mass distribution, including backgrounds from
$\gaga\rightarrow\gaga$ (dashed line extending to $\sim 130$~GeV) 
and $e\gamma\rightarrow e\gamma$ 
(heavy solid line extending to $\sim 135$~GeV) as well as the signal (peak
at 120~GeV).  The hatched histogram
shows the sum of background contributions.}
\end{center}
\end{figure}

Next we introduced resolved photon backgrounds -- see 
Section~\ref{sec:resolved} for details.  Events simulated
with Pythia are overlayed with the signal and background
events before clustering.  Significant `extra' energy
occasionally lands within the cluster cone, leading to
a reconstructed mass that is higher than the true mass.
It also partially compensates for the loss due to threshold
and clustering effects. The reconstructed mass 
distribution has a tail to high values, as shown in 
Fig.~\ref{fig:hggrecmass}.

The parametrization of the $\gaga\rightarrow\gaga$ background
was handled as before.  However, a new background was considered,
in which Compton scattering gives an energetic electron and photon
in the detector with the electron mis-identified as a photon.
This happens if the tracking device or track reconstruction fail,
or if the electron radiates most of its energy while passing 
through the beam pipe or Silicon vertex detector. As a starting 
point, we took the electron mis-identification to be $\epsega = 10^{-4}$.
This gives a background roughly as large as the background from
$\gaga\rightarrow\gaga$.  Such a small value for $\epsega$ may
require a special detector design.
\par
The distribution of reconstructed $\gaga$ mass is shown
in Fig~\ref{fig:hggrecmass}.  Besides the smooth distribution
from $\gaga\rightarrow\gaga$, which ends at $\Mgaga = 130$~GeV,
the contribution from Compton scattering is shown, extending up
to $\Mgaga = 135$~GeV.  The difference in shape can be exploited
to discriminate the two contributions.  The signal peak is
clearly visible, despite the smearing due to the resolved photon
background.  

%----------------------------------------------------------------
\section{\label{sec:radion}The role of photon-photon collisions
in exploring the scalar sector of the Randall-Sundrum model.}

The scalar sector of the Randall-Sundrum (RS) model comprises the Higgs boson
and the radion.  In general, the bare radion and bare Higgs boson mix
by virtue of a $\xi R H^\dagger H$ mixing contribution to the
visible brane Lagrangian. A non-zero value for $\xi$ is certainly allowed
by all symmetries, and so large values of $|\xi|$ are a possibility.
The phenomenology of the  Higgs boson, $\h$, and radion, $\phi$, 
mass eigenstates is strongly dependent upon $\xi$.
To fully explore and test the scalar sector phenomenology of the $\h$
and $\phi$ will require a full complement of accelerators.  Here,
we briefly sketch the crucial role of a $\gam\gam$ collider 
in verifying the scalar sector structure. This work
uses the procedures of Ref.~\cite{Dominici:2002jv} for analyzing
the RS scalar sector.   

The basic structure of the scalar sector is fixed by the four parameters,
$\mh$, $\mphi$, $\xi$ and $\gam\equiv v/\lphi$
where $v=246\gev$ and $\lphi$ is the vacuum expectation value of the
bare radion field, a new physics scale that should lie roughly
in the range $1\tev$ to $20\tev$, resulting in a range
for $\gam$ from $0.25$ down to $0.01$ (we will consider values
as high as $0.31$). A very important constraint on the
parameter space is provided by requiring that the kinetic
energy terms of the diagonalized Lagrangian not be tachyonic.
Algebraically, this reduces to the requirement that
$Z^2\equiv 1+6\xi\gam^2(1-6\xi)>0$.  This constraint implies
a maximum value of $|\xi|$ of order $1/\gam$.
A second crucial constraint on the $\mh,\mphi,\xi,\gam$ parameter
space comes from the requirement that inversion to the bare
parameters $m_{h_0}^2$ and $m_{\phi_0}^2$ be possible.  This requirement
implies that the minimum value of $|\mphi-\mh|$ 
must increase with increasing $|\xi|$, with $\mphi=\mh$ only possible
for zero mixing, $\xi=0$.

For our discussion, it will
be convenient to replace $\mphi$ by another
observable, namely $g_{ZZ\h}^2$ which will be very well
determined in $\epem\to Z\h$ production at a future $\epem$
linear collider (LC), independently of any information regarding the $\phi$.
The interesting question is then whether the $\gam\gam$ collider
can play an important role given other possible measurements
of $\h$ and $\phi$ production at the Large Hadron Collider (LHC) and the LC.

Let us write $h_0=d \h +c \phi$ and $\phi_0=a\phi+b\h$, where
$a,b,c,d$ are functions of the four parameters used to specify the
scalar sector.  All $f\anti f$ and $VV$ ($V=W,Z$) couplings
of the physical $h$ [$\phi$] are given by $\gfvh=(d+\gam b)$ 
[$\gfvphi=(c+\gam a)$] times the value for a SM Higgs boson.  
The sum rule, $\gfvh^2+\gfvphi^2=R^2$, where 
$R^2=\left[1+{\gamma^2(1-6\xi)^2\over Z^2}\right]$, will play
an important role below; 
$R^2$ is clearly $\geq 1$ (and possibly substantially larger
than 1 given that small values of $Z^2$ are possible).
The $gg$ and $\gam\gam$ couplings of the $\h$ and $\phi$ receive 
anomalous contributions from their $\phi_0$
components proportional to $\gamb$ and $\gama$ as well as 
from SM particle loops proportional to $\gfvh$ and $\gfvphi$, respectively.
These anomalous couplings are very crucial to unraveling the RS scalar
sector and it is the ability of a $\gam\gam$ collider to
probe the anomalous $\gam\gam$ coupling that would be of particular value.

\begin{figure}[t]
%\vspace*{-6in}
\hspace*{-.25in}
\resizebox{1.08\textwidth}{!}{
\rotatebox{0}{\includegraphics{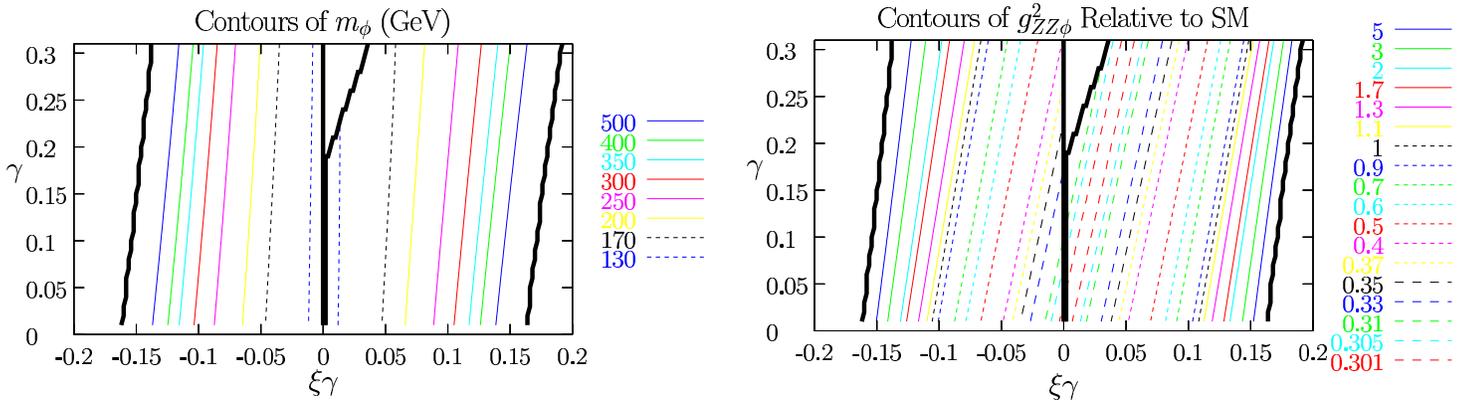}}}
%\vspace*{-2.9in}
\caption{Contours of $\mphi$ and $\gfvphi^2$
in the $(\xi\gam,\gam)$ parameter space for fixed $\mh=120\gev$
and $\gfvh^2=0.7$. Solid black contours indicate the boundaries
imposed by theoretical consistency and LEP/LEP2 direct discovery limits.
Where the inner two black boundaries at 
$\xi\gam\simeq 0$ are next to one another,
there is no actual boundary. For simplicity, in this plot we have
not displayed a 2nd set of solutions to $\mh=120\gev,\gfvh^2=0.7$
that emerge for small $|\xi\gam|$ values.
}
\label{fig:mphigfvphi}
\end{figure}

We illustrate the general strategy by focusing on just one possible
case. We imagine that the linear collider (and LHC) find
a Higgs boson of mass $\mh=120\gev$ and that the LC
determines $\gfvh^2=0.7$ (from the measurement of $g_{ZZ\h}^2$)
with very small error bars ($\lsim 3-5\%$ should be achievable, see \cite{lcvol2}).
The sum rule noted above implies that
$\gfvphi^2\geq 0.3$ (possibly substantially so) and, therefore,
if the $\phi$ is not too heavy,
the $\phi$ too will be observable in $\epem\to Z\phi$ production
at the LC and possibly also observable at the LHC. In such a case,
$\mphi$ and $\gfvphi^2$ will be well measured. Assuming that the RS
scenario is correct and that these measurements
both have very high experimental precision, 
this will be sufficient to determine
the location in $\mh,\mphi,\xi,\gam$ space up
to a two-fold ambiguity. This is illustrated in 
Fig.~\ref{fig:mphigfvphi}.\footnote{For the case being considered
it is unlikely that the $\phi$ is too heavy to be observed since a heavy
$\phi$ would probably be inconsistent with
precision electroweak constraints for $\gfvphi^2\geq 0.3$.}
One need only locate the two points (one in the $\xi<0$ region
and one in the $\xi>0$ region) at which the contours
for the measured value of $\gfvphi^2$ crosses those for the measured
value of $\mphi$. Of course, if $\gfvphi^2$ is measured with
less than perfect precision, additional ambiguity in the parameter
space location will result. It is also important to note
from Fig.~\ref{fig:mphigfvphi}
that $\mphi>\mh$ (as is generally the case if $\gfvh^2<1$) with
values of $\mphi$ close to $\mh$ only allowed when $|\xi\gam|$
is relatively small. 
However, the contours with $\mphi>300\gev$  displayed are
problematical. Given the associated $\gfvphi^2$ values, the
radion contributions to precision electroweak observables might be in
conflict with current constraints. 

There is still one further ambiguity. Namely, given just the measurements
mentioned so far, it would not be possible to say for certain which
of the two particles detected should be identified with the
eigenstate $\h$ most closely connected to the bare Higgs field
and which should be identified with the $\phi$ that derives
from the radion.

Given the inevitable errors for the $\gfvphi^2$ and $\gfvh^2$ 
measurements,\footnote{The masses $\mh$ and $\mphi$ will be
extremely well measured and errors in these quantities will not be
an issue.} the above-mentioned two-fold ambiguity, 
and the possible misidentification of $\h$ vs. $\phi$,
it will be crucial to have additional measurements to
fully determine the parameter space location.  
In this regard, we must first repeat the important fact that
all $f\anti f$ and $VV$ couplings depend only on the combinations
$\gfvh$ and $\gfvphi$. Further, the $\h$ total width is only mildly
sensitive to $\gamb$, with the consequence that 
its $f\anti f$ and $VV$ branching
ratios are only weakly dependent upon $\gamb$.   Thus, 
all rates involving only $f\anti f$ and/or $VV$ initial and final
states are almost entirely determined by the value of $\gfvh^2$,
which we have assumed to be well-measured with a value of $0.7$.
 For the case being considered where $\gfvphi^2$ is possibly as big
or bigger than $\gfvh^2$, similar statements apply to the $\phi$.
In particular, once $\mphi>2\mw$, the $WW$ and $ZZ$ branching ratios
of the $\phi$ are substantial and rather insensitive to $\gama$. 
(Total widths measured directly or indirectly would also not provide
a clear identification of $\phi$ vs. $\h$ in the case considered.)

\begin{figure}[t]
%\vspace*{-1.5in}
\hspace*{-.2in}
\resizebox{1.05\textwidth}{!}{
\rotatebox{0}{\includegraphics{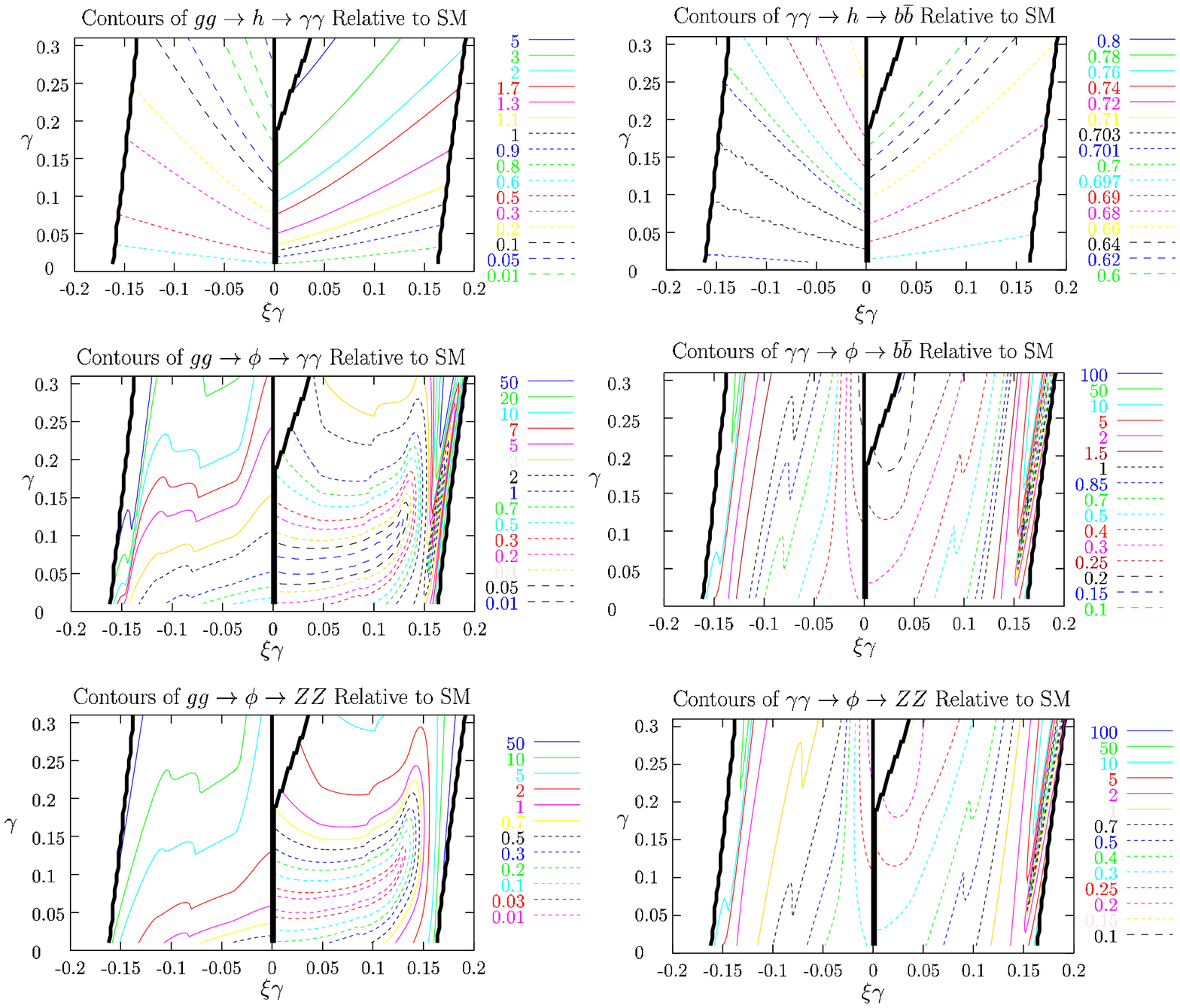}}}
%\vspace*{-2.5in}
\caption{Contours of rates (relative to a SM Higgs of the same mass)
for $gg\to \h\to \gam\gam$, $gg\to\phi\to \gam\gam$, $gg\to\phi\to ZZ$.
$\gam\gam\to\h\to b\anti b$, $\gam\gam\to \phi\to b\anti b$,
and $\gam\gam\to\phi\to ZZ$ 
in the $(\xi\gam,\gam)$ parameter space for fixed $\mh=120\gev$
and $\gfvh^2=0.7$. Solid black contours indicate the boundaries
imposed by theoretical consistency and LEP/LEP2 direct discovery limits.
Where the inner two black boundaries at $\xi\gam\simeq 0$
are next to one another,
there is no actual boundary. 
For simplicity, in this plot we have
not displayed a 2nd set of solutions to $\mh=120\gev,\gfvh^2=0.7$
that emerge for small $|\xi\gam|$ values. Note that contours of a given
ratio value for $\gam\gam\to \phi\to b\anti b$ and $\gam\gam\to\phi\to ZZ$
coincide according to the RS model.
}
\label{fig:gggagahphisummary}
\end{figure}

To clarify these ambiguities, it will be vital 
to measure $\h$ and $\phi$ production/decay channels involving $gg$ and/or
$\gam\gam$ that are sensitive to the
$gg$ and/or $\gam\gam$ anomalous couplings. There are four very useful
possibilities. By performing all four measurements it will also
be possible to fully test the structure of the RS theory, including
the anomalous $gg$ and $\gam\gam$ couplings of the $\h$ and $\phi$.
The four process are $gg\to \h\to \gam\gam$, $\gam\gam\to\h\to b\anti b$,
$gg\to \phi\to X$ ($X=\gam\gam$ for $\mphi\lsim 150\gev$
or $X=\wp\wm,ZZ$ for $\mphi\gsim 150\gev$) and $\gam\gam\to \phi \to X'$
($X'=b\anti b$ for $\mphi\lsim 160\gev$ and $X'=\wp\wm,ZZ$ for $\mphi\gsim 160\gev$.)  The ratios of these rates to the corresponding rate for
a SM Higgs boson of the same mass are shown 
in Fig.~\ref{fig:gggagahphisummary}. 
In exploring
the implications of these plots, one needs to keep in mind that
the $\gam\gam\to\phi \to b\anti b$ rates are only substantial for 
$\mphi\lsim 140\gev$, where measurements at the $\sim 3\%$
level are possible in the case of the SM Higgs~\cite{AsnerNLC}. 
For higher $\mphi$, the 
$\gam\gam\to\phi\to \wp\wm,ZZ$ rates are generally robust but experimental
studies of the accuracy with which they can be measured are not available.

We note that there is a certain complementarity between
the $gg\to\phi$ rates and the $\gam\gam\to \phi$ rates.
In particular, where the $gg\to\phi\to\gam\gam$ and $ZZ$ rates
are too suppressed to be detectable ($0<\gam\xi<0.15$, $\gam\div 0.05-0.1$),
the $\gam\gam\to \phi\to b\anti b$ and $ZZ$ rates remain adequate for
reasonably precise measurement.

Fig.~\ref{fig:gggagahphisummary} shows 
that the $\gam\gam\to \h \to b\anti b$ rate is only very modestly
sensitive to parameter space location, whereas the 
$\gam\gam\to\phi\to b\anti b$ and $\gam\gam\to\phi\to ZZ$ rates are 
very sensitive to the choice of $(\xi\gam,\gam)$. 
The $\h$ eigenstate would then be identified
as that consistent with the predicted rate for one of the two possible
choices for parameters consistent with the observed
$\mh,\mphi,\gfvh^2,\gfvphi^2$ values.  The $gg\to\h\to\gam\gam$ rate
would then provide an extremely good determination of the precise location
in $(\xi\gam,\gam)$ parameter space (given the roughly `orthogonal'
nature of the rate contours as compared to $\mphi$ and $\gfvphi^2$
contours).  The corresponding $\phi$ rates would also serve
this purpose while at the same time verifying the typically
much larger anomalous $\gam\gam$ and $gg$ couplings of the $\phi$
(determined by $\gama$ which is much bigger than $\gamb$ over most
of parameter space).

Indeed, if we look at the $\gam\gam\to\phi\to b\anti b$ contours
at small $|\xi\gam|$ where $\mphi\lsim 140\gev$ 
(see Fig.~\ref{fig:mphigfvphi}), so that $BR(\phi\to b\anti b)$
is substantial, we see in Fig.~\ref{fig:gggagahphisummary}
dramatic and for the most part
accurately measurable (where the rate is a substantial
fraction of the SM rate) variation of the rate as a function of $\gam$,
allowing an accurate determination of $\gam$ that can be checked against
the determination from $\h$ measurements. 
If we look along the $\mphi=300\gev$ contours of Fig.~\ref{fig:mphigfvphi},
$BR(\phi\to ZZ)$ will be big and Fig.~\ref{fig:gggagahphisummary} shows
strong variation of the $gg\to\phi\to ZZ$ and $\gam\gam\to\phi\to ZZ$
rates, which rates should be accurately measurable where suppression
relative to the SM is not large.  Note that all these rates are
very different for $\xi\gam<0$ as compared to $\xi\gam>0$ and
will strongly distinguish between the $\xi\gam<0$ and $\xi\gam>0$
contour crossings in Fig.~\ref{fig:mphigfvphi}
that give the same $\mphi$ and $\gfvphi^2$ values.

In fact, by using all these measurements,
a model independent determination of the
anomalous couplings of the $\h$ and $\phi$ is possible.  We define
\beq
R_{sgg}\equiv{ g_{s gg}^2(\mbox{with anomaly})\over  g_{s gg}^2(\mbox{without anomaly})}\,,
\quad\mbox{and}\quad
R_{s\gam\gam}\equiv { g_{s \gam\gam}^2(\mbox{with anomaly})\over 
 g_{s \gam\gam}^2(\mbox{without anomaly})}\,,
\label{ratiodefs}
\eeq 
for $s=\h$ and $s=\phi$. The procedure for determining $R_{sgg}$ and $R_{s\gam\gam}$ in a model-independent manner is the following. (We assume
for purposes of illustration that the $\phi$ is light enough
that its branching ratio to $\wp\wm,ZZ$ final states is small.)

First, obtain
$g_{ZZs}^2$ (defined relative to the SM prediction
at $\mhsm=m_s$) 
from $\sigma(\epem\to Z s)$ (inclusive recoil technique).
Next, determine $BR(s\to b\anti b)=\sigma(\epem\to Zs\to Zb\anti b)/
\sigma(\epem\to Zs)$. 
Then, $ g_{s\gam\gam}^2(\mbox{from experiment})
=\sigma(\gam\gam\to s\to b\anti b)/BR(s\to b\anti b)$ (note the need
for the $\gam\gam$ collider measurement).
One would then compute
\beq
R_{s\gam\gam}\equiv
{ g_{s\gam\gam}^2(\mbox{from experiment})
\over  g_{\hsm \gam\gam}^2(\mbox{as computed for $\mhsm=m_s$})
\times g_{ZZs}^2(\mbox{from experiment})}
\label{rgamgamdef}
\eeq

To determine $ g_{sgg}^2$ experimentally requires one more step.
We must compute $\sigma(gg\to s\to \gam\gam)/BR(s\to \gam\gam)$.
To obtain $BR(s\to \gam\gam)$,
we need a measurement of $\Gamma_s^{\rm tot}$.
Given such a measurement, we then compute
\beq
BR(s\to\gam\gam)={\Gamma(s\to\gam\gam)(\mbox{computed from the
experimentally determined 
$ g_{s\gam\gam}^2$})\over \Gamma_s^{\rm tot}(\mbox{from experiment})}\,,
\eeq 
where the above experimental determination of $ g_{s\gam\gam}^2$
is employed and the experimental techniques outlined
in \cite{lcvol2} are employed for $\Gamma_s^{\rm tot}$.
The ratio analogous to Eq.~(\ref{rgamgamdef})
for the $gg$ coupling is then 
\beq
R_{sgg}\equiv
{ g_{sgg}^2(\mbox{from experiment})
\over  g_{\hsm gg}^2(\mbox{as computed for $\mhsm=m_s$})
\times g_{ZZs}^2(\mbox{from experiment})}\,.
\label{rggdef}
\eeq
For a light SM Higgs boson, the various cross sections
and branching ratios needed for the $s\gam\gam$ coupling
can be determined with errors of order a few percent. 
A careful study is needed to assess the errors in the case
of the $\h$ and $\phi$ of the RS scalar sector.

\begin{figure}[t]
%\vspace*{-5.2in}
\hspace*{-.2in}
\resizebox{1.05\textwidth}{!}{
\rotatebox{0}{\includegraphics{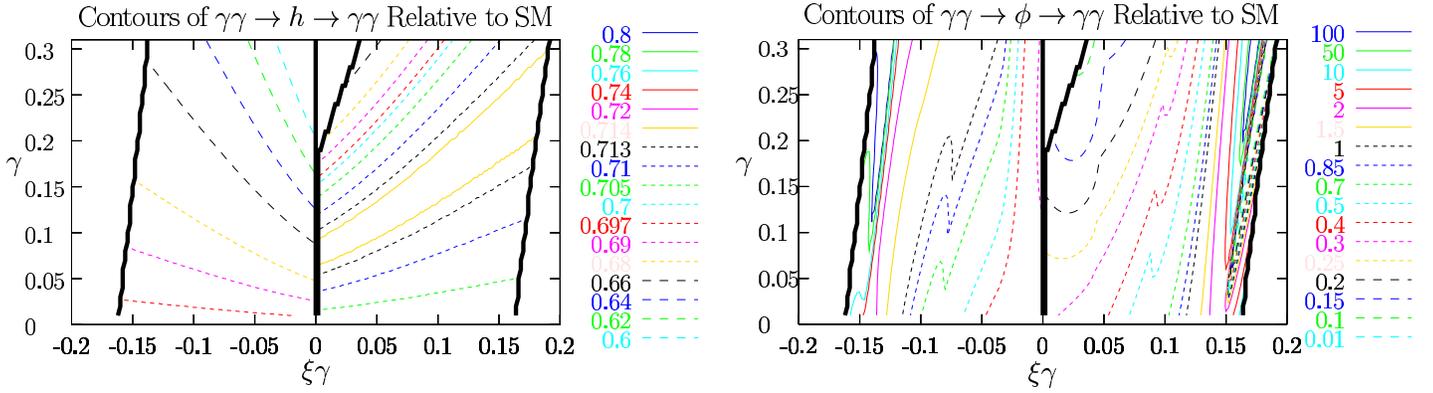}}}
%\vspace*{-2.7in}
\caption{Contours of rates (relative to a SM Higgs of the same mass)
for $\gam\gam\to \h\to \gam\gam$ and $\gam\gam\to\phi\to \gam\gam$
in the $(\xi\gam,\gam)$ parameter space for fixed $\mh=120\gev$
and $\gfvh^2=0.7$, with conditions and conventions
as in Fig.~\ref{fig:gggagahphisummary}.
}
\label{fig:gagagagahphisummary}
\end{figure}

Finally, it is interesting to see what roles the $\gam\gam\to \h \to \gam\gam$
and $\gam\gam\to\phi\to\gam\gam$ rate measurements might play.
(We see from the previous section that determination
of the $\gam\gam\to\hsm\to\gam\gam$ rate would 
be possible with $\sim 10\%$ accuracy after a few years
of operation for $\mhsm\sim 120\gev$.)  We plot contours of these
rates (relative to the corresponding $\hsm$ rate) in Fig.~\ref{fig:gagagagahphisummary}.
As expected, with $\gfvh^2=0.7$ fixed, the $\gam\gam\to\h\to\gam\gam$
rate is roughly $0.7$ times the SM value, but with up to $\sim 15\%$
variation in either direction depending upon location in $(\xi\gam,\gam)$
parameter space.  Thus, measurement of this rate with the expected
accuracy would be an interesting check and constraint on the RS model.
The $\gam\gam\to\phi\to\gam\gam$ rate shows a lot
of variation in the small $|\xi\gamma|$ region where $\mphi\lsim 140\gev$ 
(see Fig.~\ref{fig:mphigfvphi}), {\it i.e.} for $\mphi$
such that $BR(\phi\to\gam\gam)$
would be large enough that the rate might be measurable.
However, for many such parameter points, the suppression relative to
the SM comparison rate is sufficiently substantial that the accuracy
of the measurment would be relatively poor. Still, if other aspects
of the RS model are verified, this would be a very interesting final check
on the model. In particular, the ratio of the 
$\gam\gam\to\phi\to\gam\gam$ rate relative to the SM comparison
differs substantially from the corresponding ratios for $\gam\gam\to\phi\to b\anti b$ and $ZZ$ (as plotted in Fig.~\ref{fig:gggagahphisummary})
 at any given $(\xi\gam,\gam)$ parameter location.

%----------------------------------------------------------------
\section{\label{sec:charginos}Chargino pair production in 
photon-photon collisions}

%{\it Talk presented by ...}

We study the potential for determining the chargino masses and mixing
angles in two-photon collisions at a future linear collider.  These
parameters determine the fundamental SUSY parameters: $M_2$, $\mu$,
$\tan\beta$, and the CP-violating phase $\cos \Phi_{\mu}$.  

At an $e^+e^-$ collider, 
the chargino pair production cross sections are sensitive to
the mixing angles via the chargino couplings to the $Z$ boson.  Thus,
by combining mass and cross section measurements from $e^+e^-$, one
can extract the fundamental parameters of the chargino sector 
\cite{GudiCharginos}.  
The situation is complicated, however, by the fact that 
the chargino production cross section in $e^+e^-$ collisions 
receives radiative corrections of up to 10\% that depend on additional 
SUSY parameters \cite{eecharginoRCs}.

\begin{figure}
\resizebox{\textwidth}{!}{
\rotatebox{270}{\includegraphics{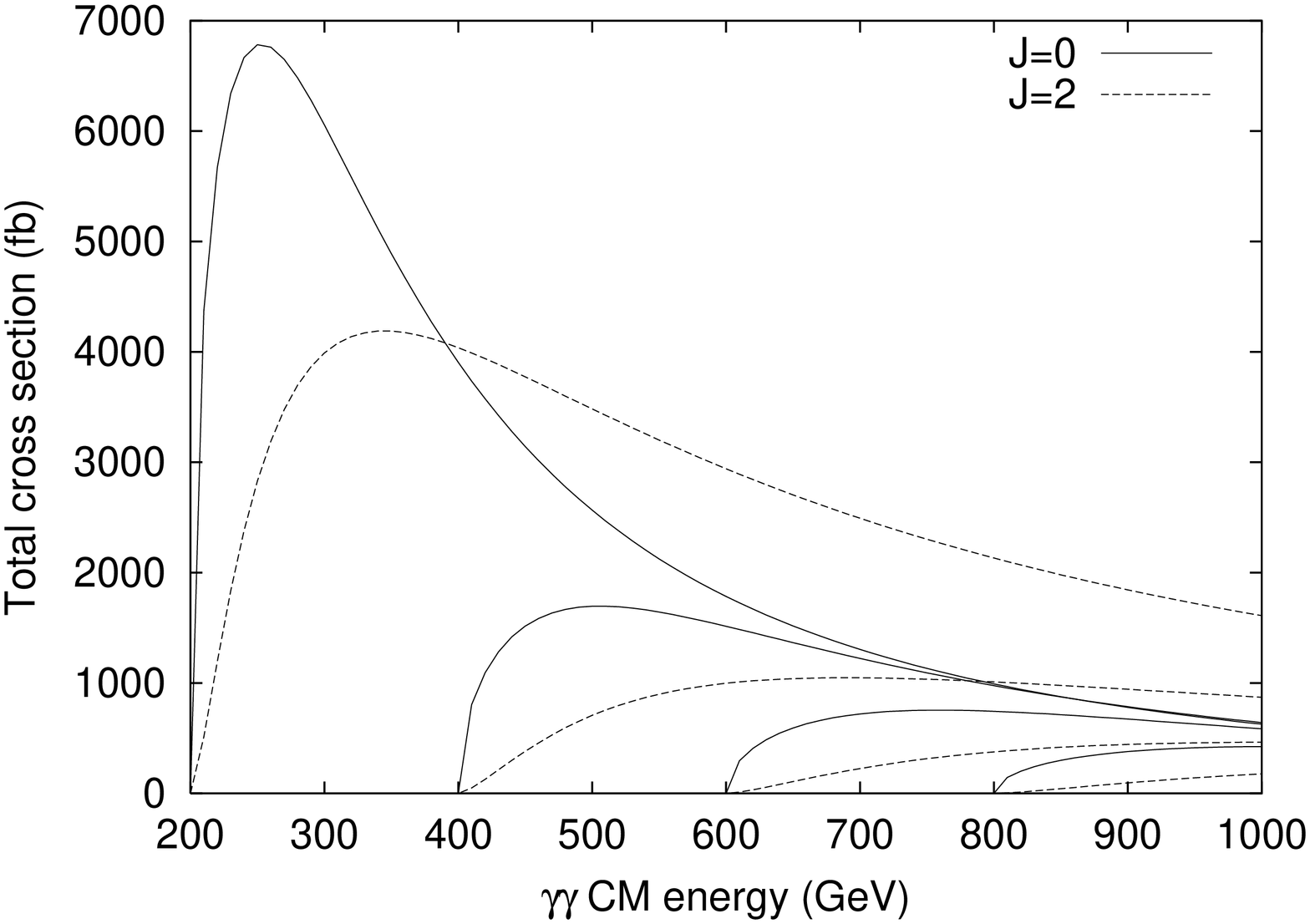}}
\rotatebox{270}{\includegraphics{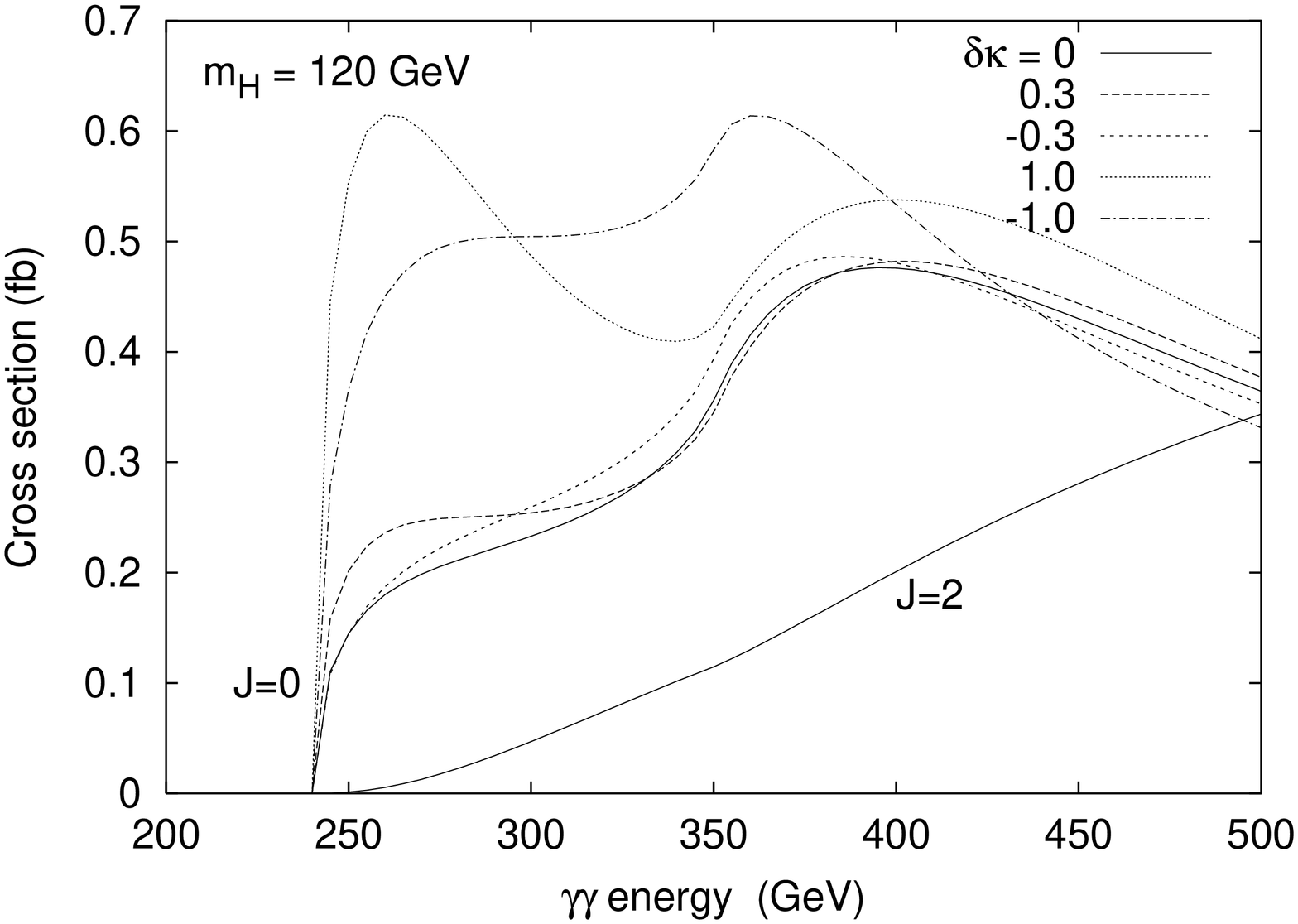}}}
\caption{(left) Parton-level cross sections in fb for chargino pair production 
in two-photon collisions as a function of the two-photon 
center-of-mass energy, for total photon helicity $J=0$ (solid curves) and 2
(dashed curves). (right) Parton-level cross section for $\gamma\gamma \to HH$ as a function
of the $\gamma\gamma$ center-of-mass energy, for $J=0$ and $J=2$.  Shown
also are the effects of varying the trilinear coupling.}
\label{fig:charginoxsec}
\end{figure}

At a $\gamma\gamma$ collider, unfortunately, there is no sensitivity
to the chargino mixing angles in the tree-level production, because
the photon-chargino couplings depend only on the electric charge.
\footnote{The decay of charginos provides some sensitivity to the mixing angles.}
However, chargino pair production at a photon collider can still be used
to measure the chargino mass.  In addition, the parameter independence of the
production cross section allows one to study the chargino decay dynamics in a 
theoretically clean way and offers sensitivity, e.g., to $M_1$ and the 
sneutrino mass through the branching ratio and forward-backward asymmetry 
in the decay $\tilde \chi_1^+ \to \tilde \chi_1^0 e^+ \nu_e$ \cite{Mayer}.
Beyond the tree level, radiative corrections
due to quark and squark loops can modify the cross section
in $\gamma\gamma$ collisions by up to 3\% \cite{Zhou}.
Of course, radiative corrections to the chargino masses \cite{charginomassRCs}
are relevant in either collider environment.
Fig.~\ref{fig:charginoxsec} shows the polarized parton-level chargino 
production cross sections \cite{Koike}.

\begin{figure}
\resizebox{.5\textwidth}{!}{
\rotatebox{0}{\includegraphics{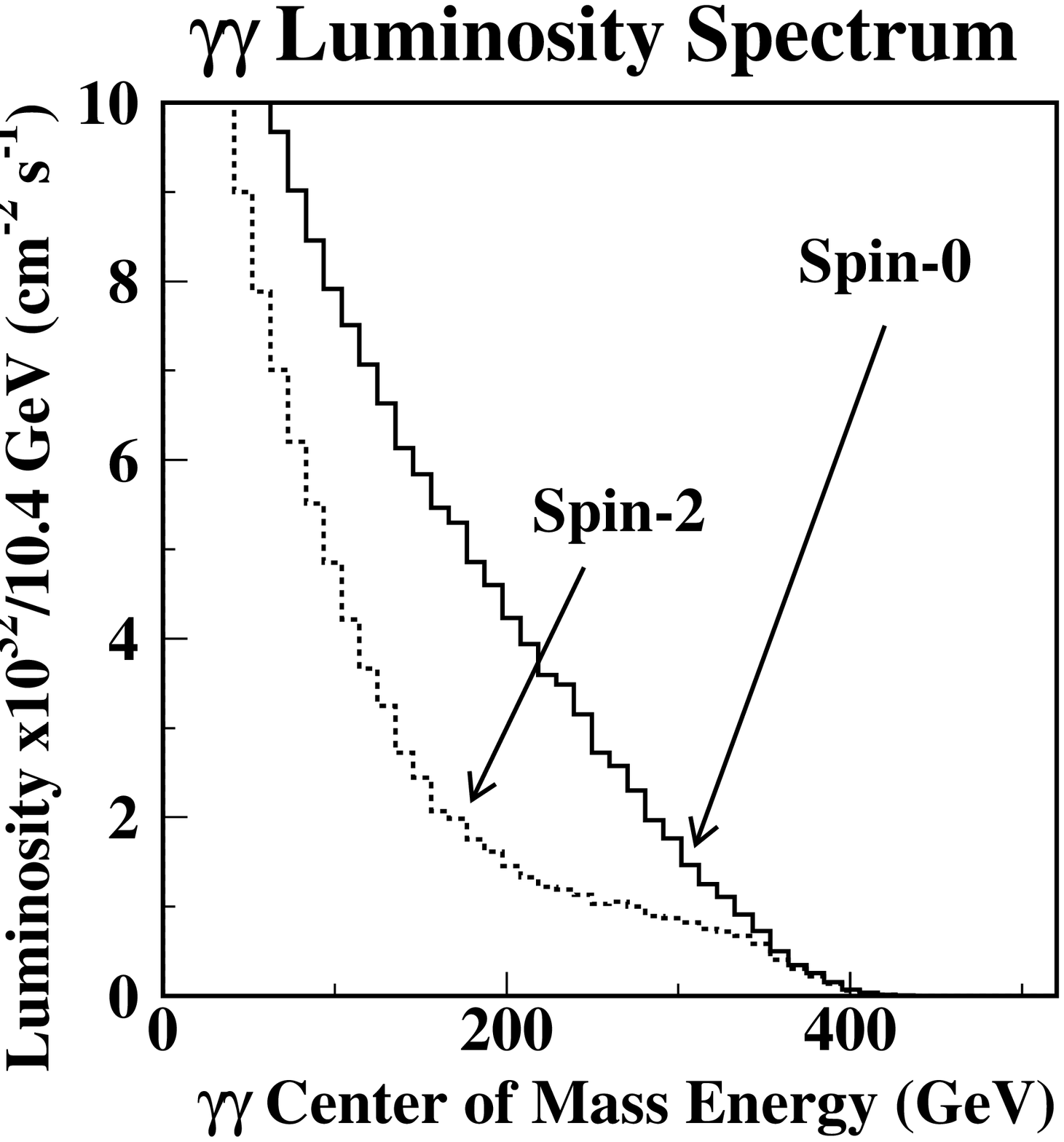}}
\rotatebox{0}{\includegraphics{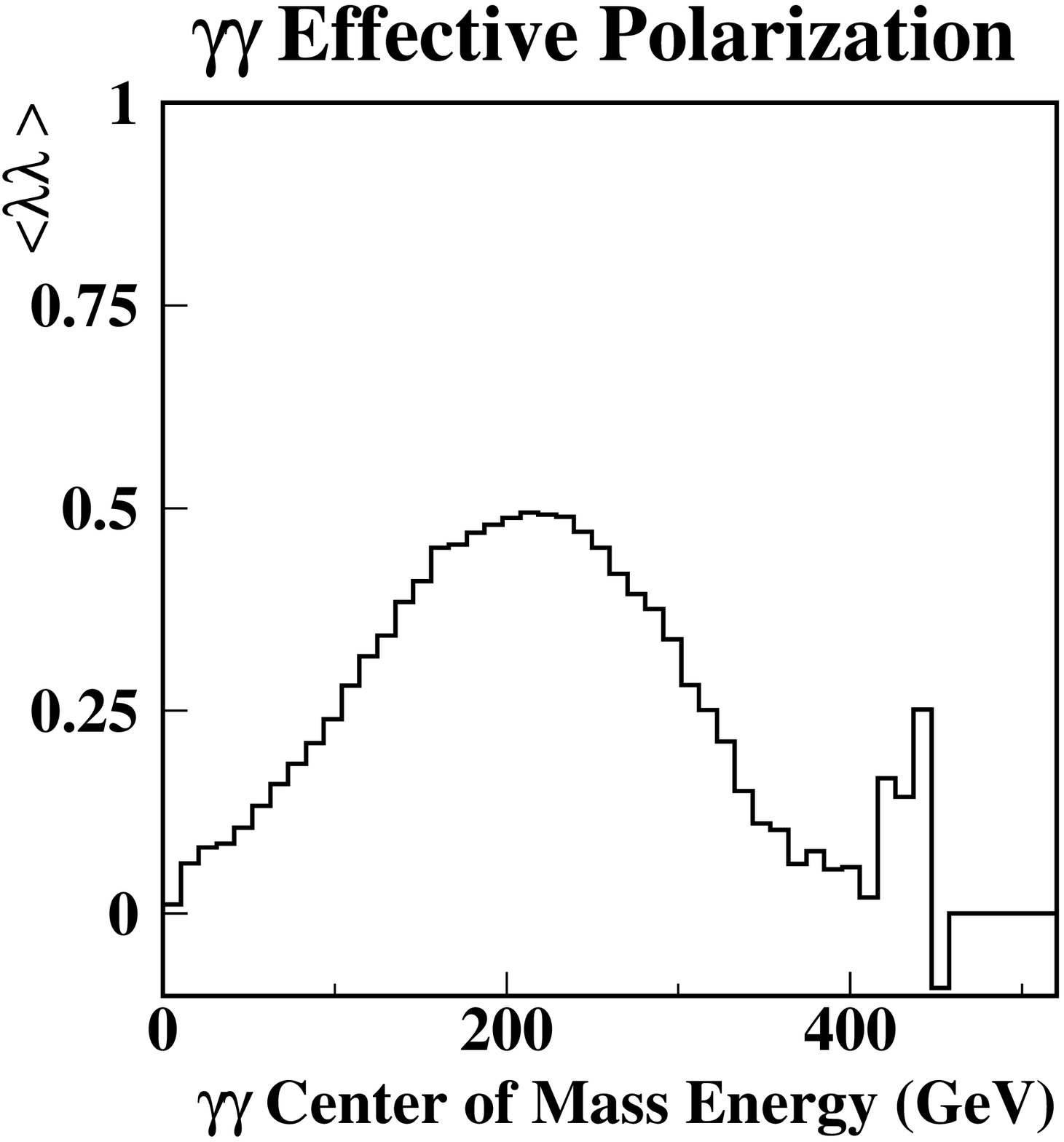}}
}\resizebox{.5\textwidth}{!}{
\rotatebox{0}{\includegraphics{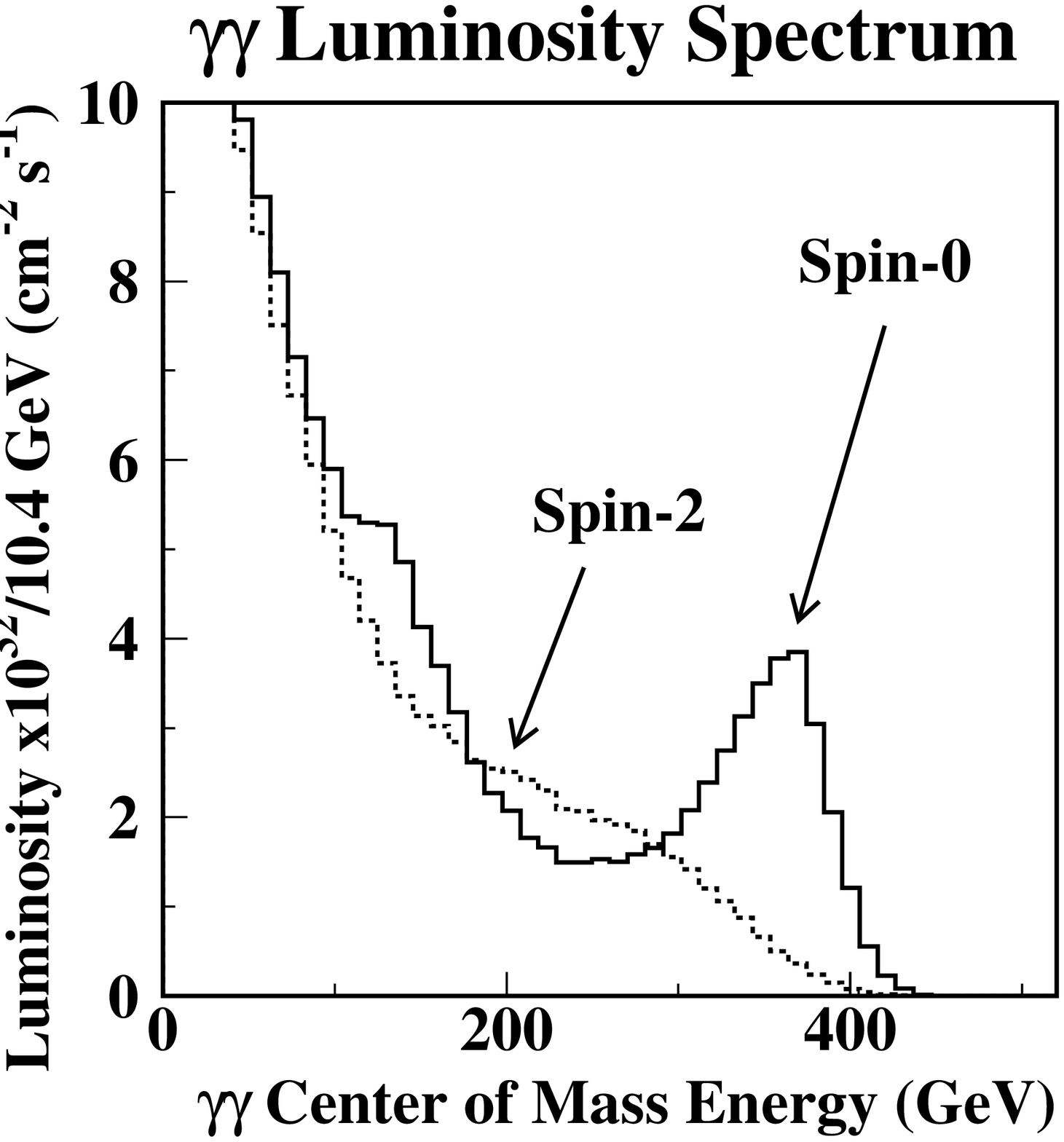}}
\rotatebox{0}{\includegraphics{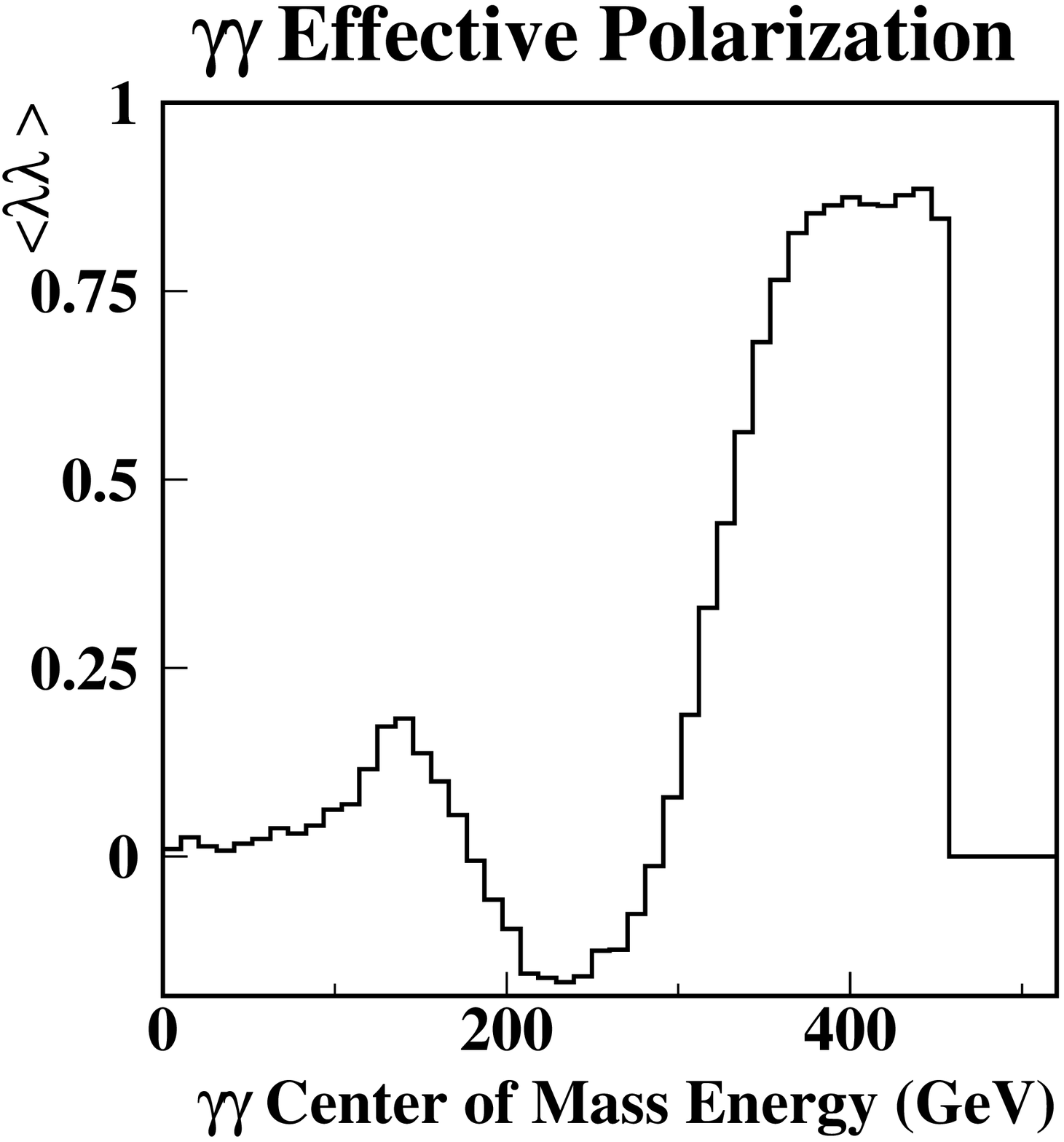}}}
%\figurebox{22pc}{15pc}{} % to have a box alone
\caption[0]{Luminosity, in units $fb^{-1}/10.4$~GeV, for a $10^7$~sec year and
associated expectation value for the product of photon polarizations, $\vev{\lam\lam'}$, are plotted for $\sqrt s_{ee}=500$~GeV
($x=4.5$ for $1.054~\mu$m laser wavelength),
assuming 80\% electron beam polarizations, for polarization
orientation cases type-I (left) and type-II (right). }
\label{fig:cainlums}
\end{figure}

Our study incorporates a ``Pantaleo'' final focus~\cite{ffs}
optimized for $\gamma\gamma$ collisions~\cite{ggffs} and realistic two-photon spectra based
on the most probable available laser technology~\cite{laser}.
We use the laser and electron beam parameters given in
Table~\ref{tab:beampar} and the CAIN~\cite{cainref}
Monte Carlo program
to obtain
the luminosity and polarization spectra plotted in Fig.~\ref{fig:cainlums}.
The CAIN simulation of the interaction between 
the laser photons and 
the primary electron beam, and the Compton scattered photons and the spent
electron beam, includes the effects due to
beamstrahlung, secondary collisions and other non-linear 
effects. These effects provide large corrections to the
naive luminosity distributions first considered in Ref.~\cite{Ginzburg:1983vm,Ginzburg:1984yr}.
The $J=0$ cross section turns on quickly, since it
goes like $\beta$ at threshold, while the $J=2$ cross section goes like
$\beta^3$.  In $e^+e^-$, the cross section goes like $\beta$ at threshold.

The estimated
event yield is given in Table~\ref{tab:charginos}, assuming a chargino mass of $m_{\tilde \chi_1^+} = 150$ GeV, along with the
cross section and luminosity described above.
\begin{table}
\begin{tabular}{rccc}
\hline
\hline
 & $\int \mathcal{L}_{th}$ (fb$^{-1}/10^7$ s) & $\sigma$ (fb) & Event yield \\
\hline
Spin-0 & 30 & 3000 & 90,000 \\
Spin-2 & 5  & 1000 & 5,000 \\
$e^+e^-$ & 160 & 300 & 48,000 \\
\hline \hline
\end{tabular}
\caption{Comparison of the integrated luminosity above threshold 
($\int \mathcal{L}_{th}$), the
``average'' cross section ($\sigma$), and the event yield per Snowmass
year of $10^7$ sec for pair production of charginos of mass 150 GeV
in $\gamma\gamma$ and $e^+e^-$ collisions.}
\label{tab:charginos}
\end{table}
The chargino yield is a factor of two greater in two-photon collisions
than in $e^+e^-$.
Because the final state is the same, we expect the signal reconstruction
efficiency to be similar to that in $e^+e^-$.
Thus, it may possible to determine the chargino masses more precisely in
$\gamma\gamma$ collisions. Further study is required since the $\gam\gam$
luminosity spectrum, shown in Fig.~\ref{fig:cliclums} and 
Fig.~\ref{fig:cainlums}, does not have a sharp end point. Luminosity beyond the
naive kinematic limit is due to multiple Compton interactions. A sharper end
point can be obtained by reducing the laser power and thus the probability for
multiple Compton interactions however the total luminosity is also reduced.
The optimal laser power for 
end point mass measurements has not been determined.

This is a report on a work in 
progress.  Future reports will include full background and detector
simulation.

%----------------------------------------------------------------
\section{\label{sec:trilinear}Measuring the trilinear coupling of the 
Standard Model Higgs boson}

%{\it Talk presented by ...}

We examine the potential for determining the trilinear coupling of
the Standard Model Higgs boson in two-photon collisions at a future
linear collider.
The trilinear Higgs boson coupling can be measured in $e^+e^-$ 
collisions \cite{eeHHZ,Gay}. 
There is also some sensitivity to
the trilinear Higgs coupling at the LHC \cite{HHLHC}.
This is a direct test
of the shape of the Higgs potential.
Accuracies of approximately 20\% on the $e^+e^-\to ZHH$ 
cross section can be obtained
for $M_H$ in the range of 120~GeV to 140~GeV with 
1000~$fb^{-1}$ at $\rtsee=500$~GeV. A neural network improves the sensitivity to 13\%
for $M_H$=120~GeV. The sensitivity 
is diluted to 22\% by additional amplitudes which are insensitive to the
trilinear coupling.

We use the cross sections for SM Higgs boson pair production in polarized
two-photon collisions computed in Ref.~\cite{JikiaHH}.  The trilinear Higgs
boson coupling enters the diagram containing a virtual $s$-channel Higgs
boson, which contributes only to the $J=0$ amplitude.  To evaluate the 
sensitivity of the cross section to the trilinear Higgs coupling, we
introduce an anomalous trilinear Higgs coupling in a gauge-invariant 
way, following Ref.~\cite{JikiaHH}:
\begin{equation}
        \delta \mathcal{L}_{\rm Higgs} = 
        - \frac{\delta \kappa}{2} \frac{m_H^2}{v} 
        \left[ H^3 + \frac{3}{v} G^+G^- H^2 \right] + \cdots,
\end{equation}
where $v = 246$ GeV is the Higgs vacuum expectation value, $H$ is the SM
Higgs field, $G^{\pm}$ are the charged Goldstone bosons, and 
$\delta \kappa$ is the dimensionless anomalous trilinear Higgs coupling
normalized so that for $\delta \kappa = 1$, the anomalous term will cancel
the SM $H^3$ coupling.
We computed the one-loop integrals with the help of the 
LoopTools package \cite{LoopTools}.
Fig.~\ref{fig:charginoxsec} shows the parton-level cross sections for 
$\gamma\gamma \to HH$ for $J=0$ and $J=2$, for the SM (solid lines)
and including an anomalous trilinear coupling (dashed lines).
        %
%\begin{figure}
%\resizebox{.5\textwidth}{!}{
%\rotatebox{270}{\includegraphics{H3roots_bw.ps}}}
%\caption{Parton-level cross section for $\gamma\gamma \to HH$ as a function
%of the $\gamma\gamma$ center-of-mass energy, for $J=0$ and $J=2$.  Shown
%also are the effects of varying the trilinear coupling.}
%\label{fig:HHH}
%\end{figure}

To estimate the event yields, given in Table~\ref{tab:HHH},
in an NLC-like machine, we use the 
machine assumptions described in Sec.~\ref{sec:charginos}.
\begin{table}
\begin{tabular}{r|ccc|ccc}
\hline
\hline
 & \multicolumn{3}{c}{$\sqrt{s_{ee}} = 500$ GeV}&\multicolumn{3}{|c}{ $\sqrt{s_{ee}} = 800$ GeV}\\
 & $\int \mathcal{L}_{th}$ (fb$^{-1}/10^7$ s) & $\sigma$ (fb) 
& Event yield & $\int \mathcal{L}_{th}$ (fb$^{-1}/10^7$ s) & $\sigma$ (fb) 
& Event yield \\
\hline
Spin-0 & 40 & 0.3 & 13 & 120 & 0.3 & 39\\
Spin-2 & 20  & 0.1 & 1-2 & 60 & 0.2 & 1-2 \\
$e^+e^-$ & 160 & 0.2 & 32 & 250 & 0.15 & 38 \\
\hline \hline
\end{tabular}
\caption{Comparison of the integrated luminosity above threshold 
($\int \mathcal{L}_{th}$), the
``average'' cross section ($\sigma$), and the event yield per Snowmass
year of $10^7$ sec for double-Higgs production
in $\gamma\gamma$ and $e^+e^-$ collisions.  We assume $m_h = 120$ GeV.}
\label{tab:HHH}
\end{table}
The spin-0 event yield is about 3 times as large at $\sqrt{s_{ee}} = 800$ GeV.
since more of the twp-photon luminosity spectrum is above $HH$ threshold.

We estimate that
production rates for $e^+e^- \to HHZ$ and $\gamma\gamma \to HH$ at
an 800 GeV linear collider are comparable.  
In $e^+e^-$ collisions, the reconstruction efficiency of the $ZHH$ final
state is 43\% \cite{Gay}.  
We expect it to be better than this in $\gamma\gamma$ 
collisions, because of the simpler $HH$ final state.
The dominant background
in both analyses is $e^+e^- / \gamma\gamma \to WW$.  We estimate
comparable sensitivity to the cross section
per running time in $\gamma\gamma$ or
$e^+e^-$ collisions at $\rtsee=800$~GeV. As in $e^+e^-$ collisions, the 
sensitivity to the trilinear coupling my be diluted due to additional
amplitudes contributing to $\gamma\gamma \to HH$.  

This is a report on a work in 
progress.  Future reports will include full background and detector
simulation.

%----------------------------------------------------------------

\section{\label{sec:H+H-}Prospects for observing charged Higgs bosons in 
gamma-gamma collisions}

In the case of the two-doublet MSSM
Higgs sector there are five physical Higgs bosons - two CP-even, $\hl$ and 
$\hh$ (with $\mhl$ $<$ $\mhh$), one CP-odd, $\ha$, and a charged Higgs pair, 
$\hpm$.
The ability to detect $\gam\gam\to \hh,\ha,H^+H^-$ will be of
greatest importance if the heavy Higgs bosons cannot be detected either
at the LHC or in $\epem$ collisions at the LC.  In fact, there is
a significant section of parameter space in the MSSM, 
often referred to as the LHC `wedge' region, for which
this is the case. 

\begin{figure}
\resizebox{.9\textwidth}{!}{
\rotatebox{0}{\includegraphics{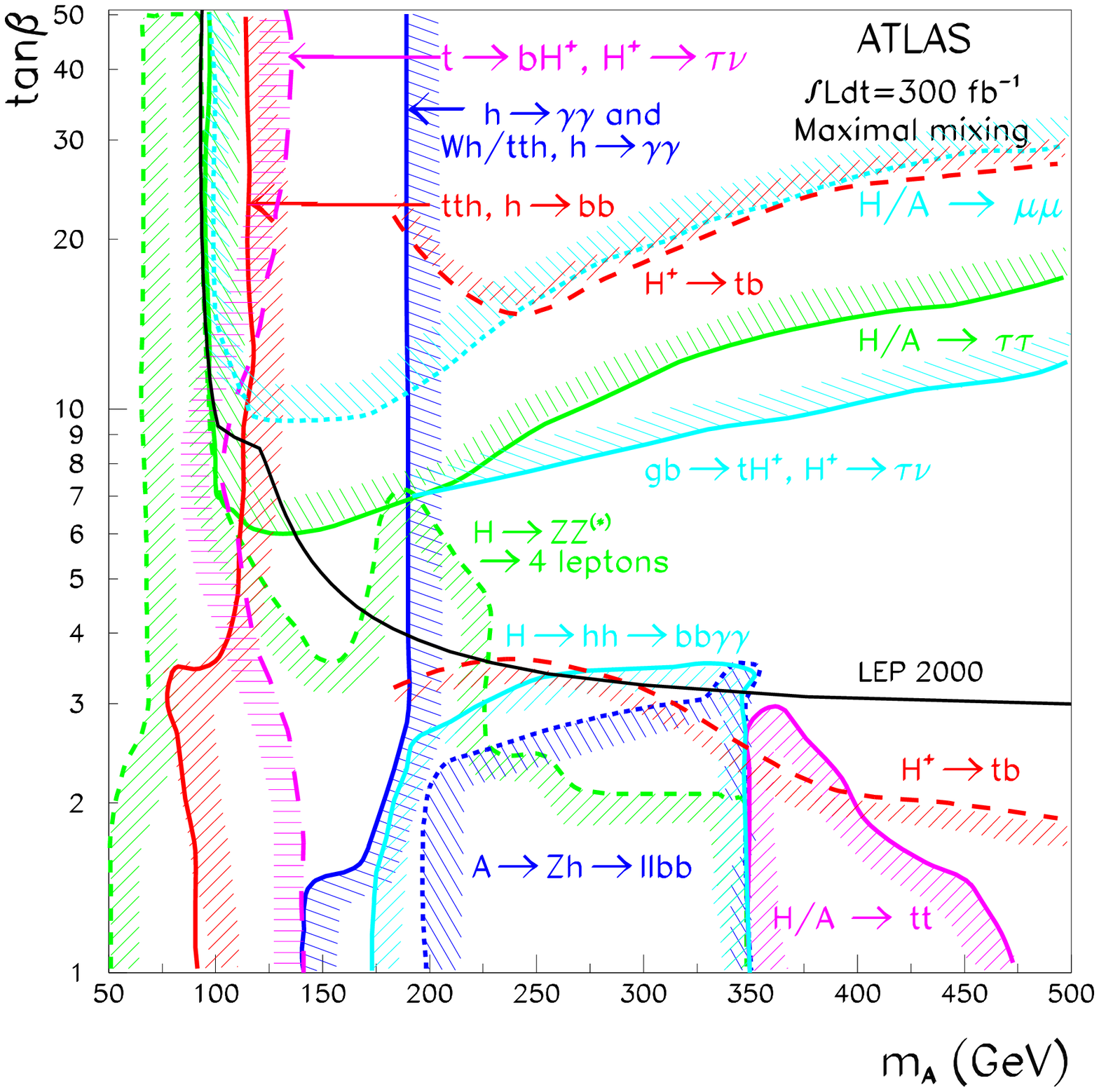}
\includegraphics{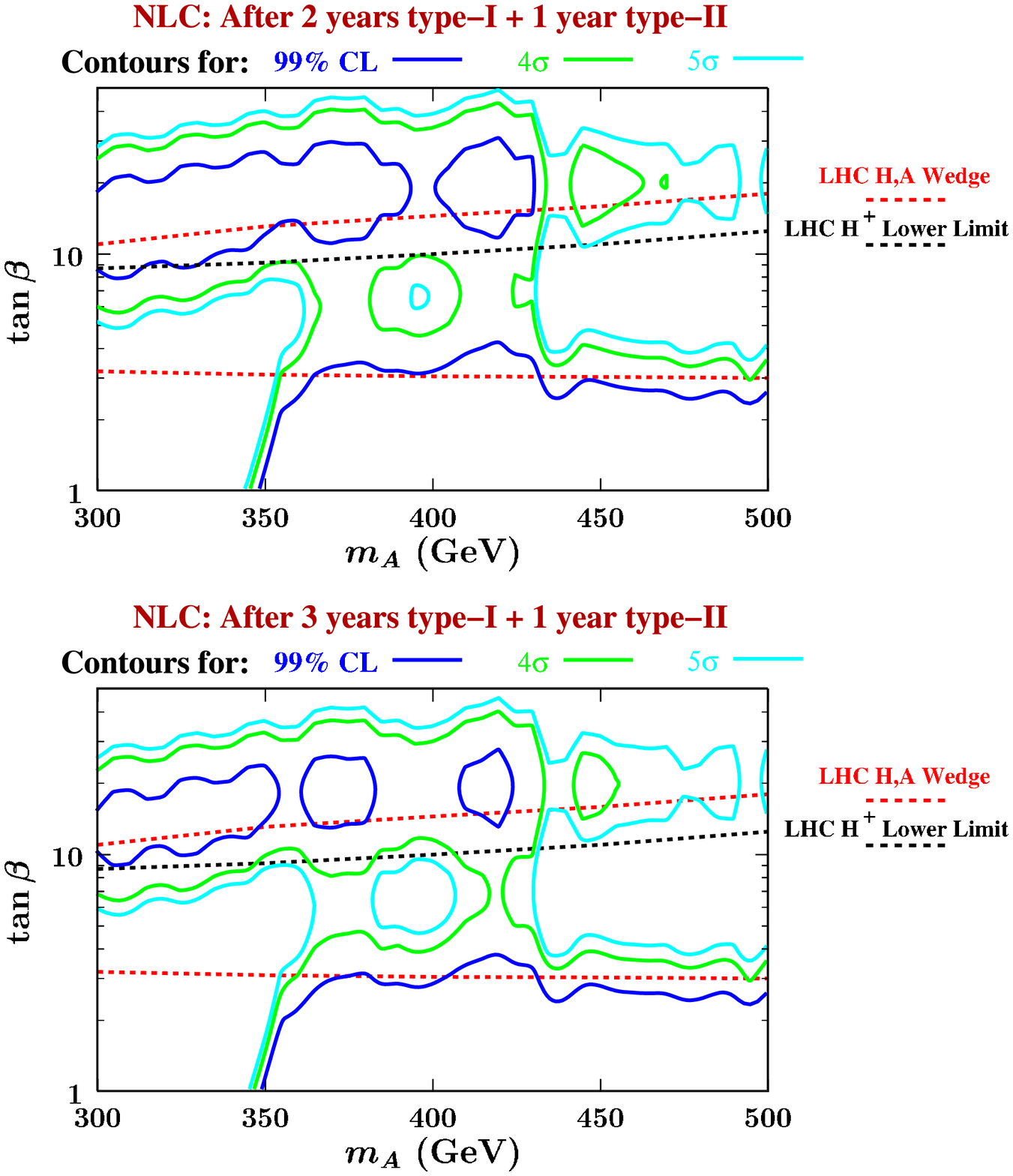}}}
\caption[0]{(left) $5\sigma$ discovery contours for MSSM Higgs boson detection
in various channels are shown in the $[\mha,\tanb]$ parameter plane, 
assuming maximal mixing and an integrated luminosity of $L=300\fbi$
for the ATLAS detector. 
 This figure is preliminary~\cite{atlasmaxmix}.
 Note that \gghh is a potential discovery mode for $\gamma\gamma$ as the $\hpm$ will be undetected at the LHC for $\mhpm>125$~GeV and moderate $\tan\beta$.
(right) The dashed lines delineate the LHC wedge region. The $4\sigma$ and $5\sigma$ discovery contours for $\hh,\ha$ in the $\gam\gam \to \hh,\ha \to b\overline b$ channel are shown. For the case where the $\hh,\ha$ are unobserved, the 99\% exclusion contour is shown. In just 3 years of operation at design luminosity the $\hh, \ha$ would be discovered (excluded) in about 2/3 (nearly all) of the LHC wedge region~\cite{AsnerNLC}.}
\label{fig:atlasmssm}
\end{figure}

We have studied the prospects for $\gam\gam\to \hh,\ha$~\cite{AsnerNLC} where
we conclude that a $\gam\gam$ collider can
provide Higgs signals for the $\hh$ and $\ha$
over a possibly crucial portion of  parameter space in which
the LHC and direct $\epem$ collisions at a LC will not be able
to detect these Higgs bosons or their $\hpm$ partners.
This is illustrated in Fig.~\ref{fig:atlasmssm}.
In this section we discuss the prospects for observing charged Higgs bosons from 
the interaction \gghh. This is a potential discovery mode for $\gamma\gamma$ as 
the $\hpm$ will be undetected at the LHC for $\mhpm>125$~GeV and moderate $\tan\beta$. 

This study assumes a $e^-e^-$ machine with $\sqrt{s_{ee}}=500\;\GeV$ and
$80\%$ electron polarization.  The CAIN Monte Carlo program~\cite{cainref}
is used to model the luminosity spectra for two configurations 
of laser polarization, type-I and type-II, 
described in Section~\ref{sec:charginos}.
%These two cases are type-I, where the laser polarization is $+1$, 
%and type-II where the laser polarization  is $-1$.  
The luminosity spectra are shown in Fig.~\ref{fig:cainlums}.  
In both of these configurations 
one Snowmass year of $10^7$ sec running would correspond to a 
total integrated luminosity of $4\times 10^{32}\;\mathrm{cm^{-2}s^{-1}}$ or 
$400\;\fb^{-1}$.

A mass range for \hpm\ from the lower experimental bound of 
$\mhpm \approx 80\;\GeVcsq$ up to the kinematic limit of 
$\mhpm = 200\;\GeVcsq$ is considered.  For $\mhpm < m_{t}$ the main decay 
mode is \htt\ as the decay modes to $W h^0$ and  $t \overline{b}$ are  
suppressed.  

\begin{figure}
\begin{center}
\includegraphics[width=.5\textwidth]{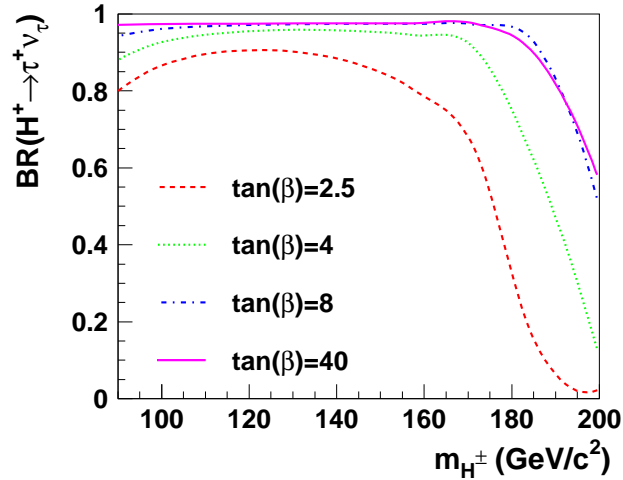}
\caption{\label{fig:httbr}
MSSM predictions of the branching ratio for the process 
\htt\ as a function of \mhpm, for various $\tanb$ values.}
\end{center}
\end{figure}

\begin{figure}\begin{center}
\resizebox{\textwidth}{!}{
\includegraphics[width=5cm]{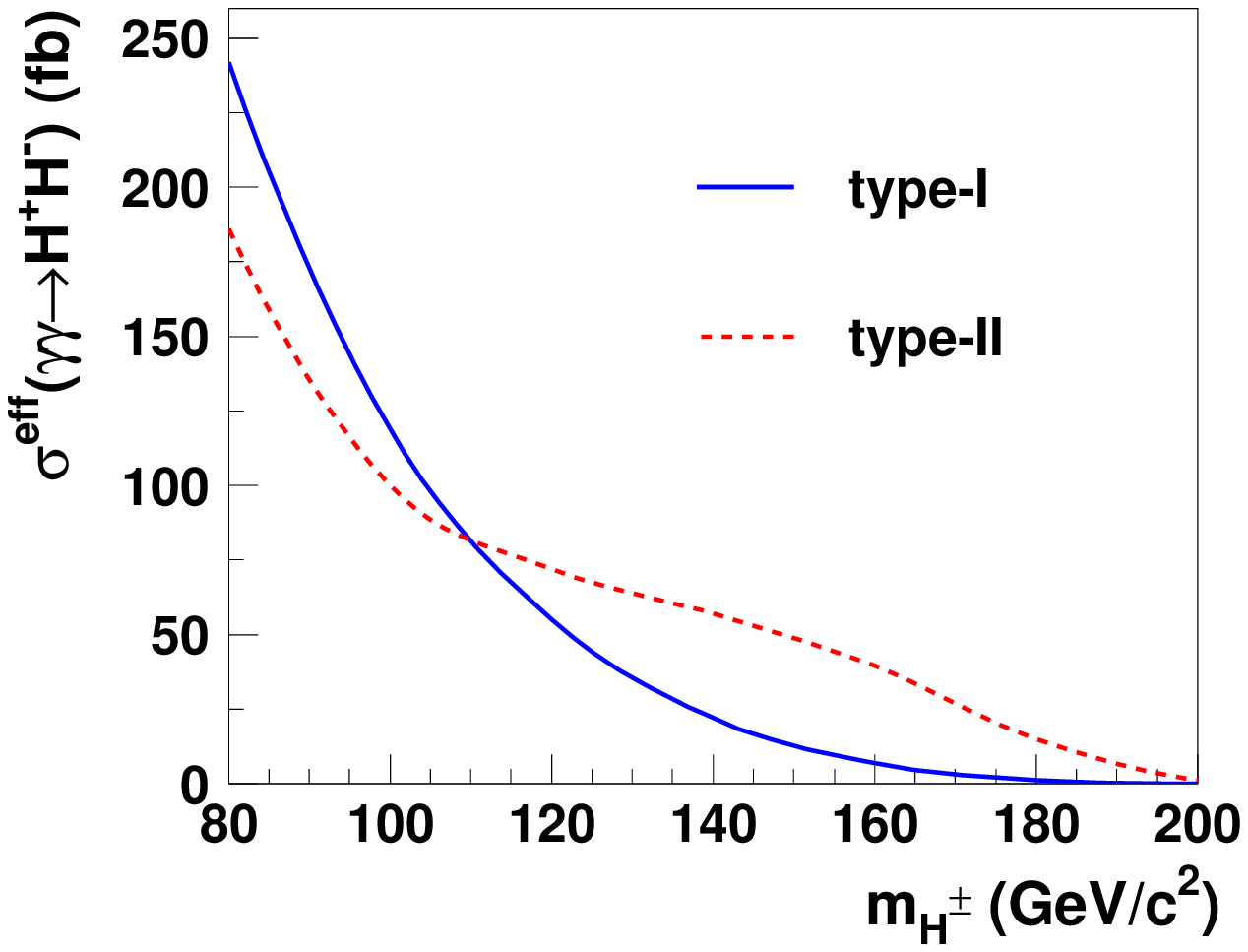}
\includegraphics[width=10cm]{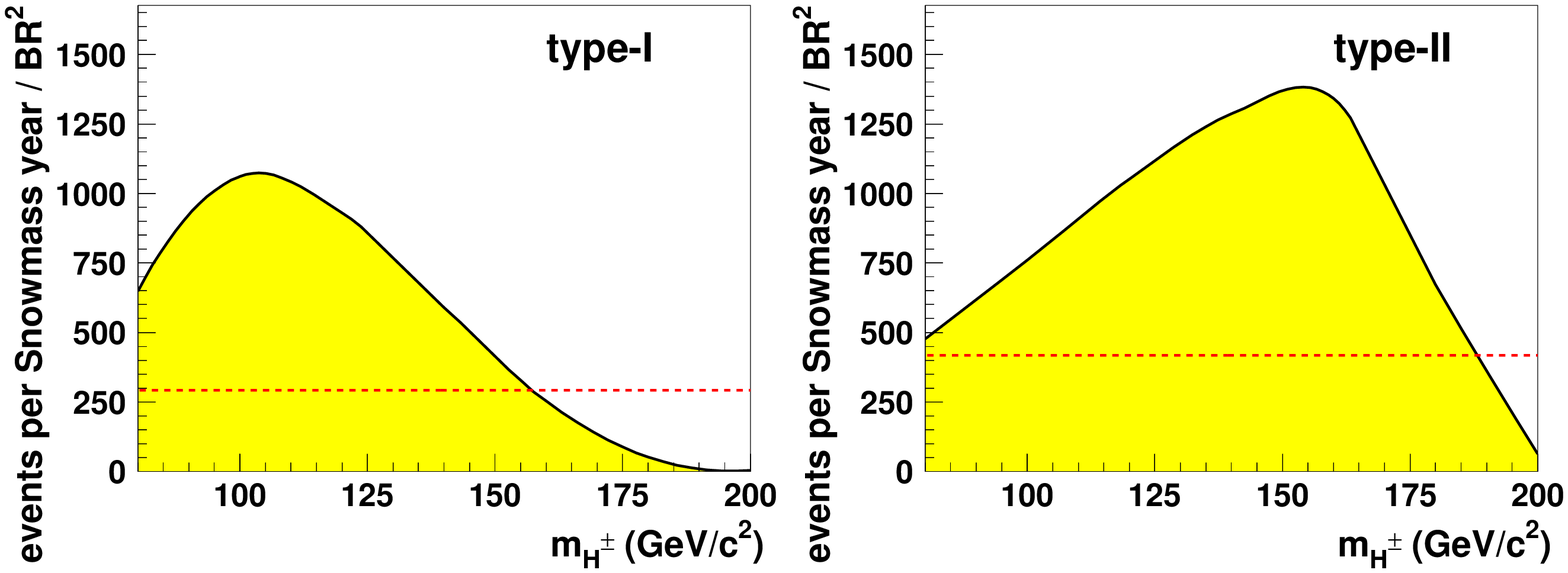}}
\caption{\label{fig:gghhxsec} On the left:
The effective cross section for \gghh\ for the two beam configurations.
The center and right plots show the number of accepted events per
 $\mathrm{BR}(\htt)^2$ per Snowmass year, as a function of \mhpm.
The dashed horizontal line shows the number of accepted background \ggww\ events.}
\end{center}\end{figure}

Fig.~\ref{fig:httbr} shows the MSSM prediction for the branching ratio of 
\htt\ as a function of \mhpm.  The values were obtained by running the HDECAY 
program~\cite{HDECAY}.
A value of $500\;\GeVcsq$ was used for the SUSY breaking mass parameters.
For most values of $\tanb$  the branching ratio is 
between $80\%$ and $98\%$ when $\mhpm < m_t$.
To allow these results to be interpreted independently of the branching ratio,
the number of events over branching ratio squared are presented.
Pandora-Pythia~\cite{pandorapythia} was used to calculate effective cross 
section for \gghh, taking into account the luminosity spectra.
The results for the two beam types is shown in Fig.~\ref{fig:gghhxsec}.

Backgrounds to the \gghh\ process come from both $\tau$ and $W$-pair 
production.  
The cross section for \ggtt\ is around $300\;\pb$ and for 
\ggwwtt\ is approximately $220\;\fb$ for the type-II beam.
For a charged Higgs mass of $140\;\GeVcsq$ these backgrounds are 
factors of approximately 5,000 and 4~times the signal cross section.

\begin{figure}\begin{center}
\includegraphics[width=5.0cm]{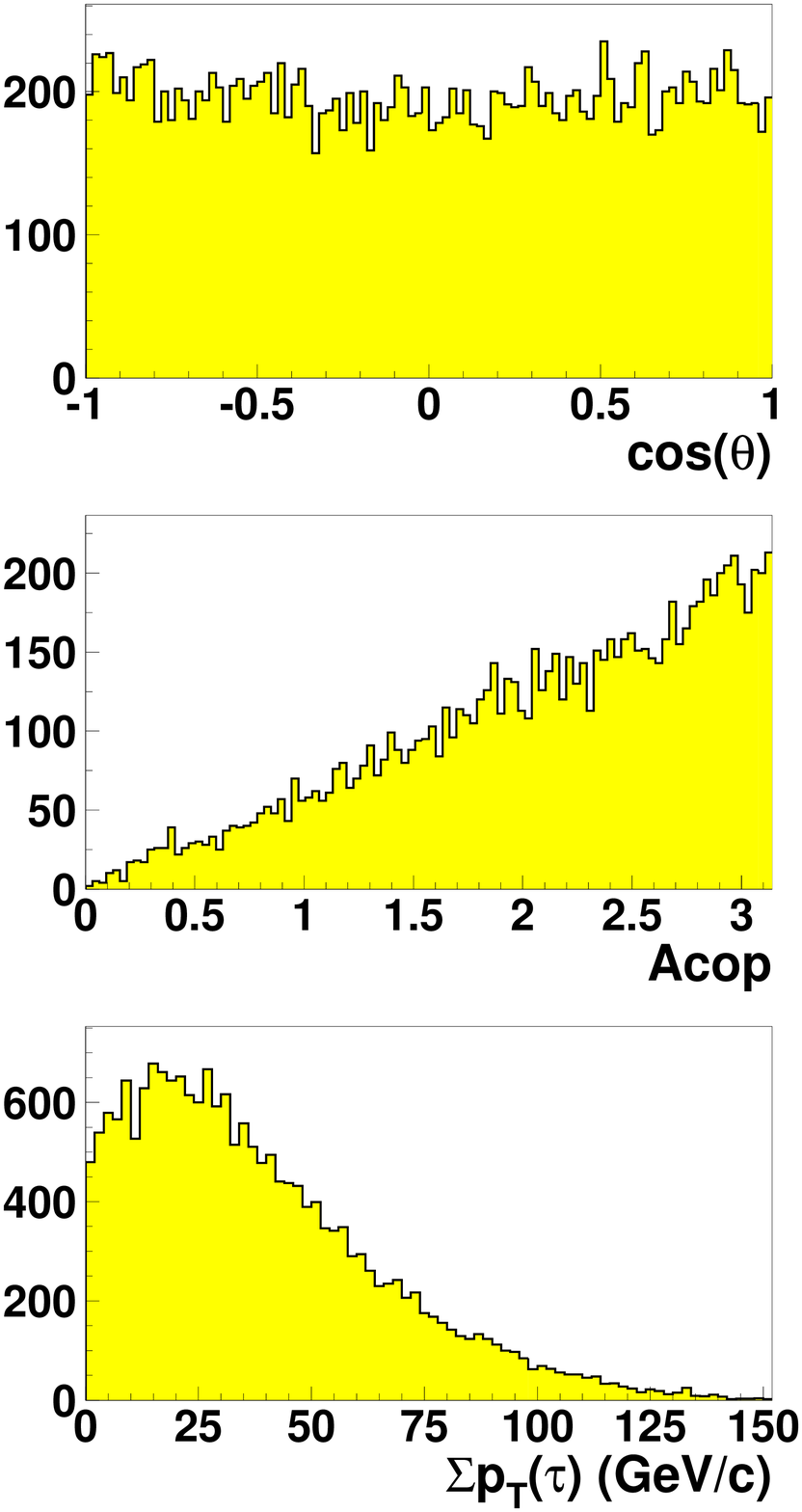}\hfil
\includegraphics[width=5.0cm]{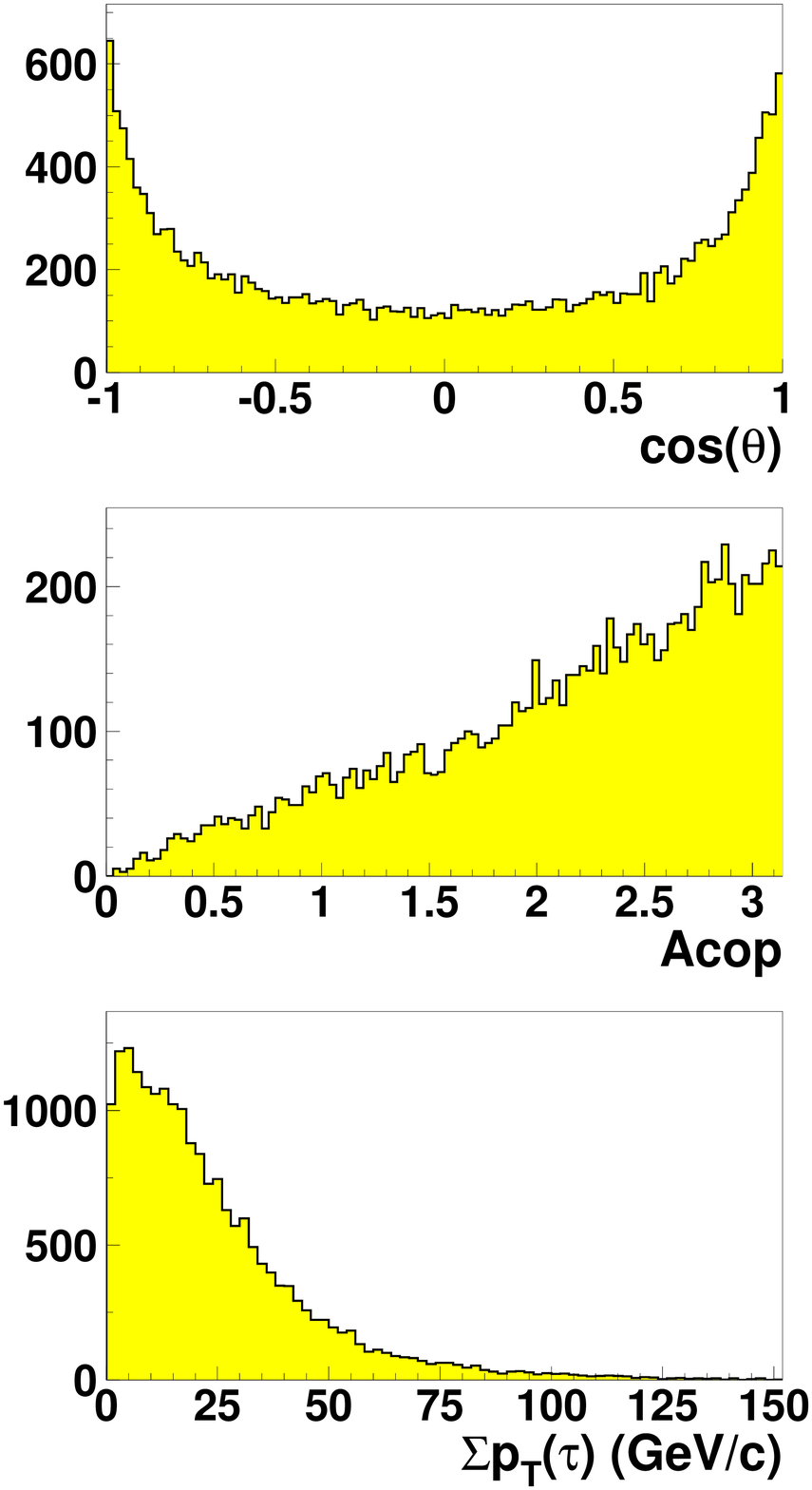}\hfil
\includegraphics[width=5.0cm]{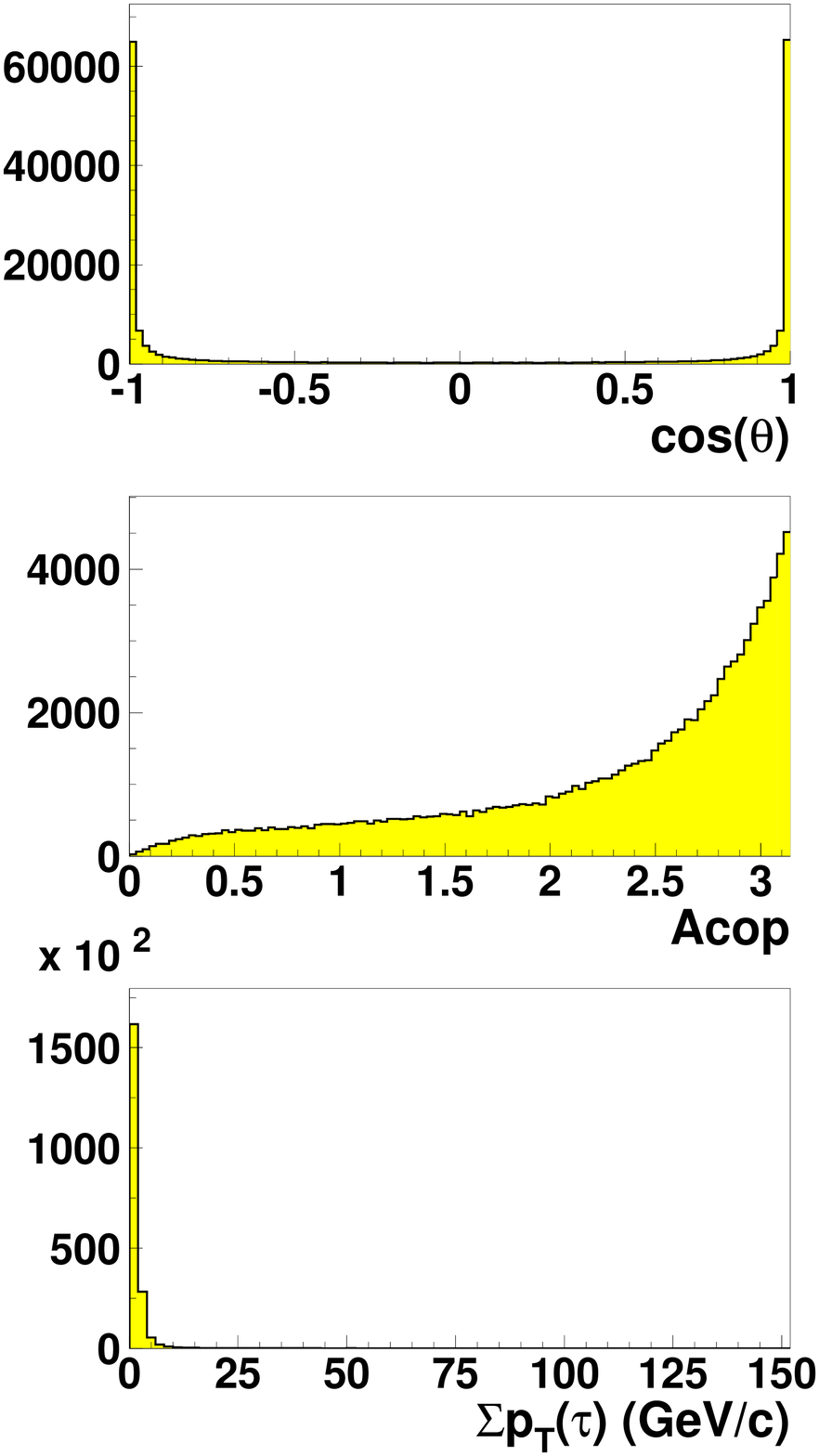}
\caption{\label{fig:gghh1}
The three observable quantities defined in the text.   
The left plots are for $H^+H^-$ events, the center plots are
for $W$-pair production and the right plots are for $\tau$-pair production.}
\end{center}\end{figure}

Three empirically reconstructable variables were used to study the signal and
background processes:
the cosine of the angle of the tau-jet in the detector, $\cos(\theta)$,
the sum of transverse momenta of all the visible particles from the tau 
decays, $\Sigma p_T (\tau)$, and
the acoplanarity of the two tau jets, 
$Acop = \cos^{-1}\left(\vec{p}_{T1}\cdot\vec{p}_{T2}/(p_{T1}p_{T2})\right)$.
We reject background with the requirements that 
$|\cos(\theta)| < 0.8$, $Acop < 3$ and 
$\Sigma p_T (\tau)/\GeV c^{-1} > 35 + 25\cdot(Acop - \frac{\pi}{2})^2$.
Fig.~\ref{fig:gghh1} shows these three quantities for the signal and 
two background processes 

These requirements reduce the \ggtt\ background by a factor of 
more than $10^{-6}$, therefore this background can be ignored.
The acceptance of the $W$-pair production background is around $0.5\%$, 
while keeping around $10\%$ of the signal for $\mhpm = 140\;\GeVcsq$.

Fig.~\ref{fig:gghhxsec} shows the number of events expected per Snowmass
year, divided by the \htt\ branching ratio squared.
%This calculation uses the cross section (also shown in 
%figure~\ref{fig:gghhxsec}), the acceptance of the listed cuts and one 
%Snowmass year integrated Luminosity of $400\;\fb^{-1}$. 
The dashed horizontal line the plots indicates the the number of \ggww\ events.

It can be seen the type-II beam configuration is the best for observing \gghh.
At a charged Higgs mass of $140\;\GeVcsq$, around 1250 events$/\mathrm{BR}(\htt)^2$ 
would be observed per Snowmass year, to be compared with about 300~\ggwwtt\ events.

For $\tanb \moresim 4$ the signal to background ratio is greater than~1 
for  $\mhpm \lesssim m_t$.  For smaller values of $\tanb$ the signal to 
background is better than~1 up to charged Higgs masses of about $160\;\GeVcsq$.
The type-I beam would also be useful for observing \gghh\ for 
$\mhpm \lesssim 130\;\GeVcsq$.

It is not possible to reconstruct a mass-type variable for the \gghh\ events,
as there are too many unknown kinematic quantities.  Studies using 
detector simulation are under way to look for observables correlated 
with \mhpm.

In conclusion, for moderate values of $\tanb$ it will be 
possible to observe \gghh\ events using a 
$500\;\GeV$  $e^-e^-$ collider for all masses up to the top quark mass.
This will probe some of the MSSM parameter space where the
$\hpm$ will be undetected at the LHC ($\mhpm>125$~GeV and 
moderate $\tan\beta$). The process $\gam\gam \to H^+H^- \to t{\overline t} b{\overline b}$ will probe the remaining region of parameter space in which
the $\hpm$ would be undetected at the LHC.

%Further work is required to determine if the mass of the \hpm\ boson can 
%be determined from these events.

%The production of charged Higgs pairs at a gamma-gamma collider
%is large, on the order of hundreds of fb.  The main backgrounds
%from W and tau pair production can be greatly reduced with a
%few kinematic cuts, leaving a fairly pure signal sample.
%An example analysis is presented for the di-tau channel, and
%the sensitivity to the charged Higgs production rate is estimated.

%Radiative corrections to $\gamma\gamma \to H^+H^-$ cross section
%from Yukawa contributions \cite{ggHHYukawa} and squark loops
%\cite{ggHHsquark}.

%----------------------------------------------------------------
\section{\label{sec:resolved} Resolved photon backgrounds}

\begin{figure}\begin{center}
\resizebox{\textwidth}{!}{
\includegraphics{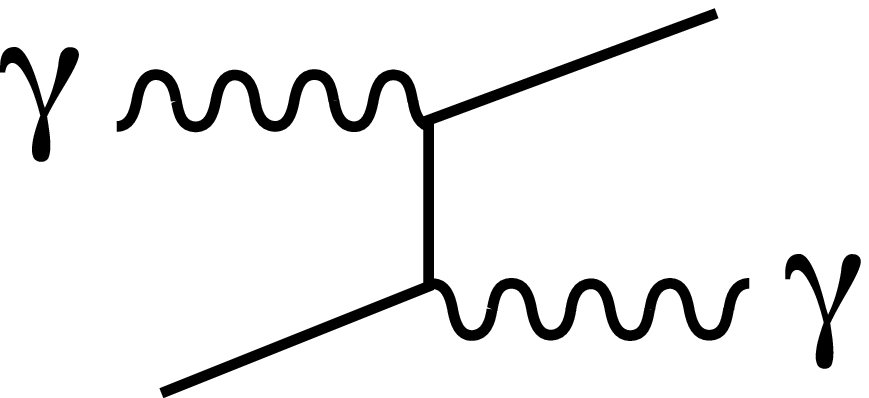}
\hskip 1cm
\includegraphics{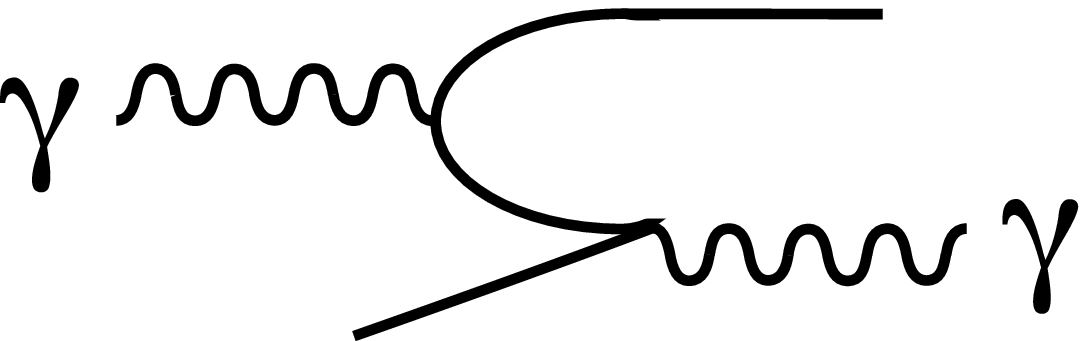}
\hskip 1cm
\includegraphics{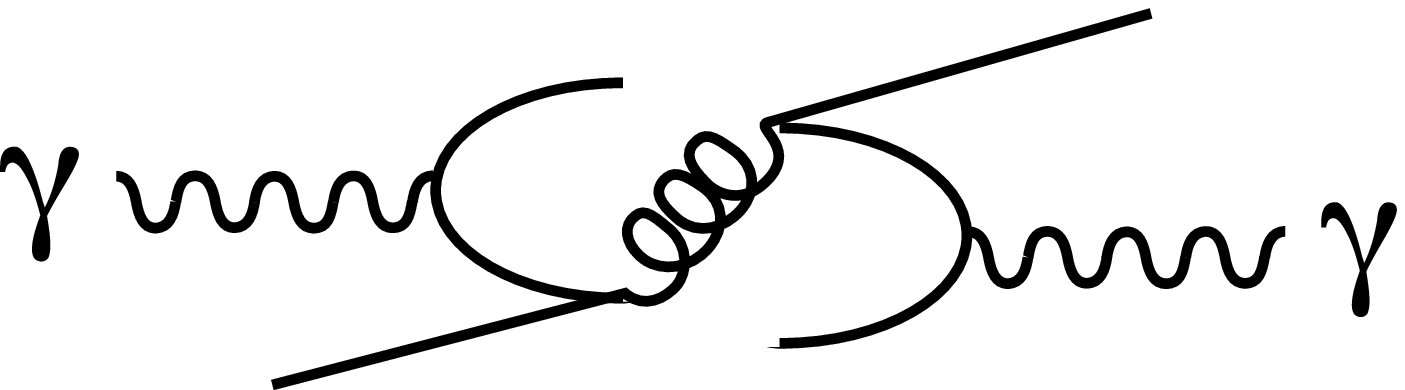}}
\caption{\label{fig:resphotons}Photon-photon Interactions. 
(left) Direct interactions involve only electroweak coupling.
(center) Once resolved process where one photon probes the parton structure of
the other photon, similar to deep inelastic scattering. (right) Twice resolved
process where the partons of each photon interact, similar to a $\rho-\rho$ collisions.}
\end{center}\end{figure}

%{\it Talk presented by ...}
The photon is the gauge boson of QED. The photon couples, via virtual charged
fermion pairs, into the electroweak and strong interactions. Photon-photon
interactions can be classified into three types of processes, illustrated
in Fig.~\ref{fig:resphotons}.
Direct interactions involve only electroweak coupling, the
once resolved process where one photon probes the parton structure of
the other photon, similar to deep inelastic scattering, and the twice resolved
process where the partons of each photon interact, similar to a $\rho-\rho$ collisions.
We use the Pythia~\cite{pythiajetset} Monte Carlo program to simulate the 
resolved photon cross sections which are potential backgrounds to {\em all}
other two-photon physics processes. These backgrounds, usually referred to as
$\gamma\gamma \to {\rm hadrons}$, are also a concern at $e^+e^-$ and have
been studied in detail~\cite{teslagghh}. However, at a $\gamma\gamma$ collider
the high energy Compton photons provide an additional and dominant source
of $\gam\gam \to {\rm hadron}$ background.
Nominally, we use the default settings for Pythia with the exception of 
the parameters listed in 
Table~\ref{tab:butterworthsettings}~\cite{butterworth}.
\begin{table}
\begin{center}
\begin{tabular}{lc}
\hline
\hline
Parameter Setting & Explanation\\
         MSTP(14)=21               & Direct gamma-gamma\\
         MSTP(14)=24               & Twice resolved photon\\
         MSTP(14)=4                & Once resolved photon\\
        PARP(67)=4   & fixes some problems with the initial \\
        PARP(71)=4   & and final state QCD radiation\\
        CKIN(3)=1.0                & ptmin  for hard 2->2 interactions\\
        PARP(81)=1.0               & ptmin for final state interactions\\

        MSTP(81)=1                 & Multiple interactions turned on (default)\\
        MSTP(82)=1                 & Structure of multiple interactions (default)\\
        PARP(90)=0.0               & Removes energy scaling\\
        PARP(89)=1.0               & Rendered useless by papr(90)=0.0\\
\hline
\hline
\end{tabular}
\end{center}
\caption{\label{tab:butterworthsettings} Pythia parameter settings for resolved photon processes.}
\end{table}

\begin{figure}\begin{center}
\resizebox{\textwidth}{!}{
\includegraphics{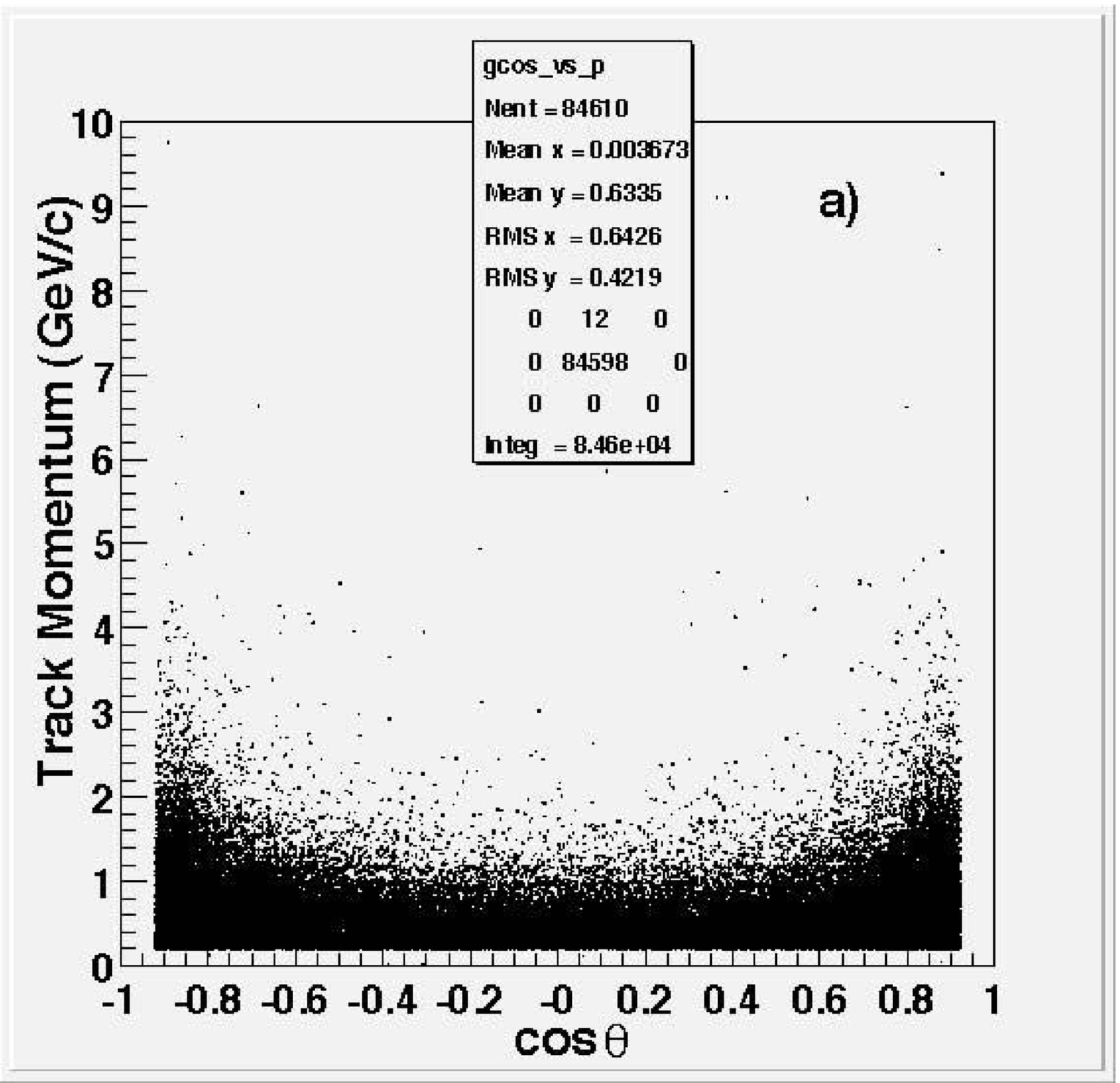}
\includegraphics{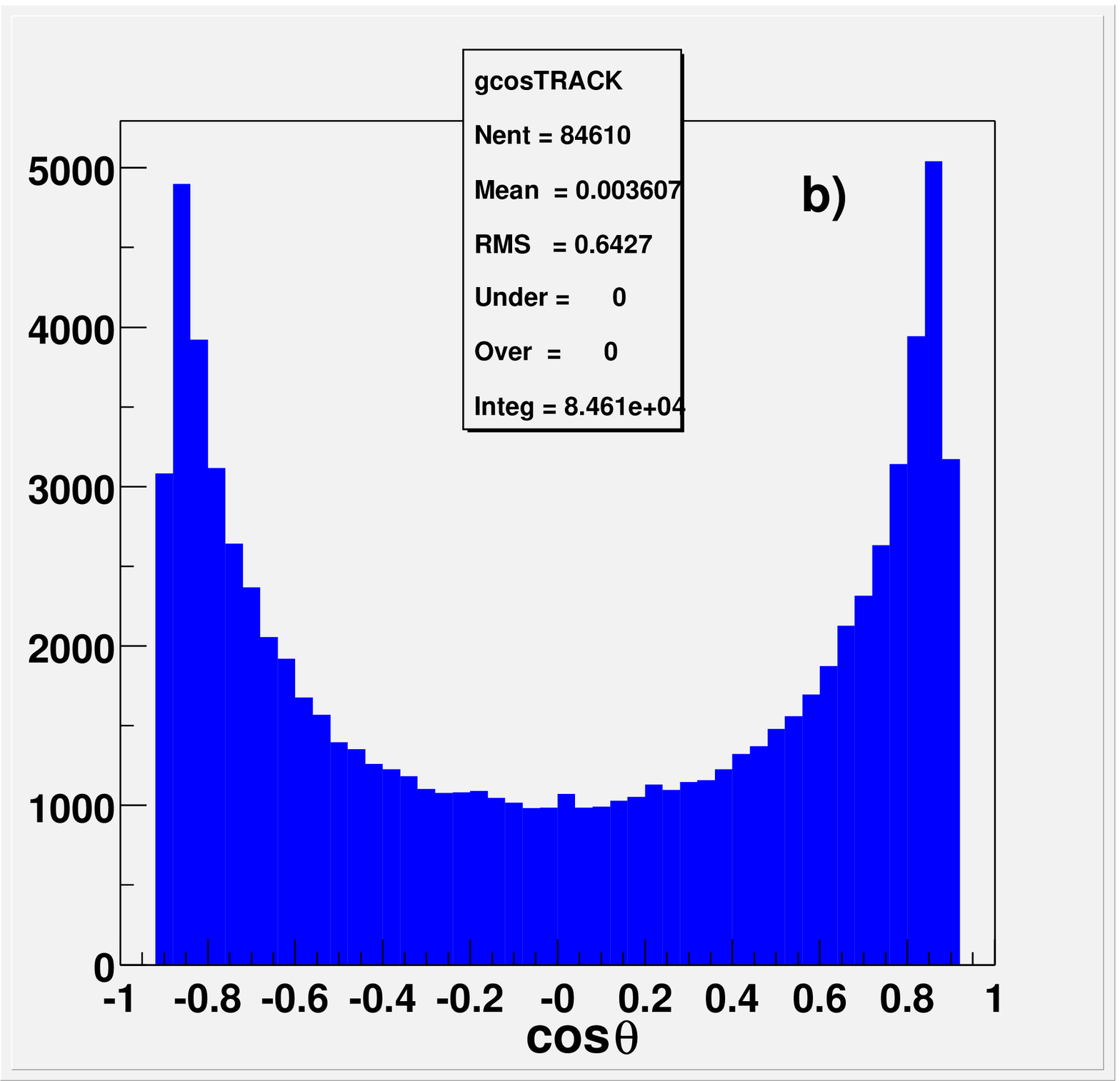}
\includegraphics{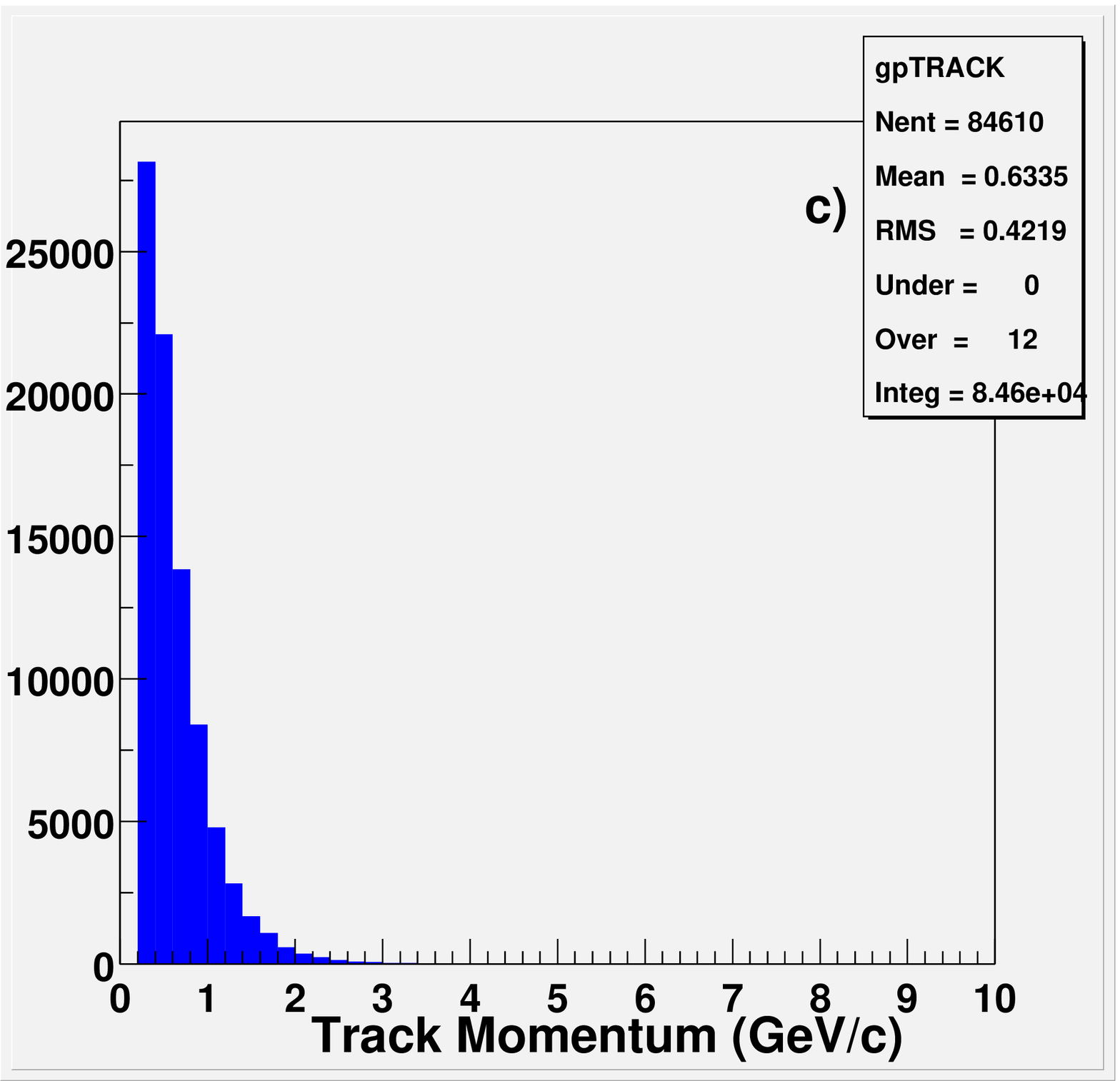}}
\resizebox{\textwidth}{!}{
\includegraphics{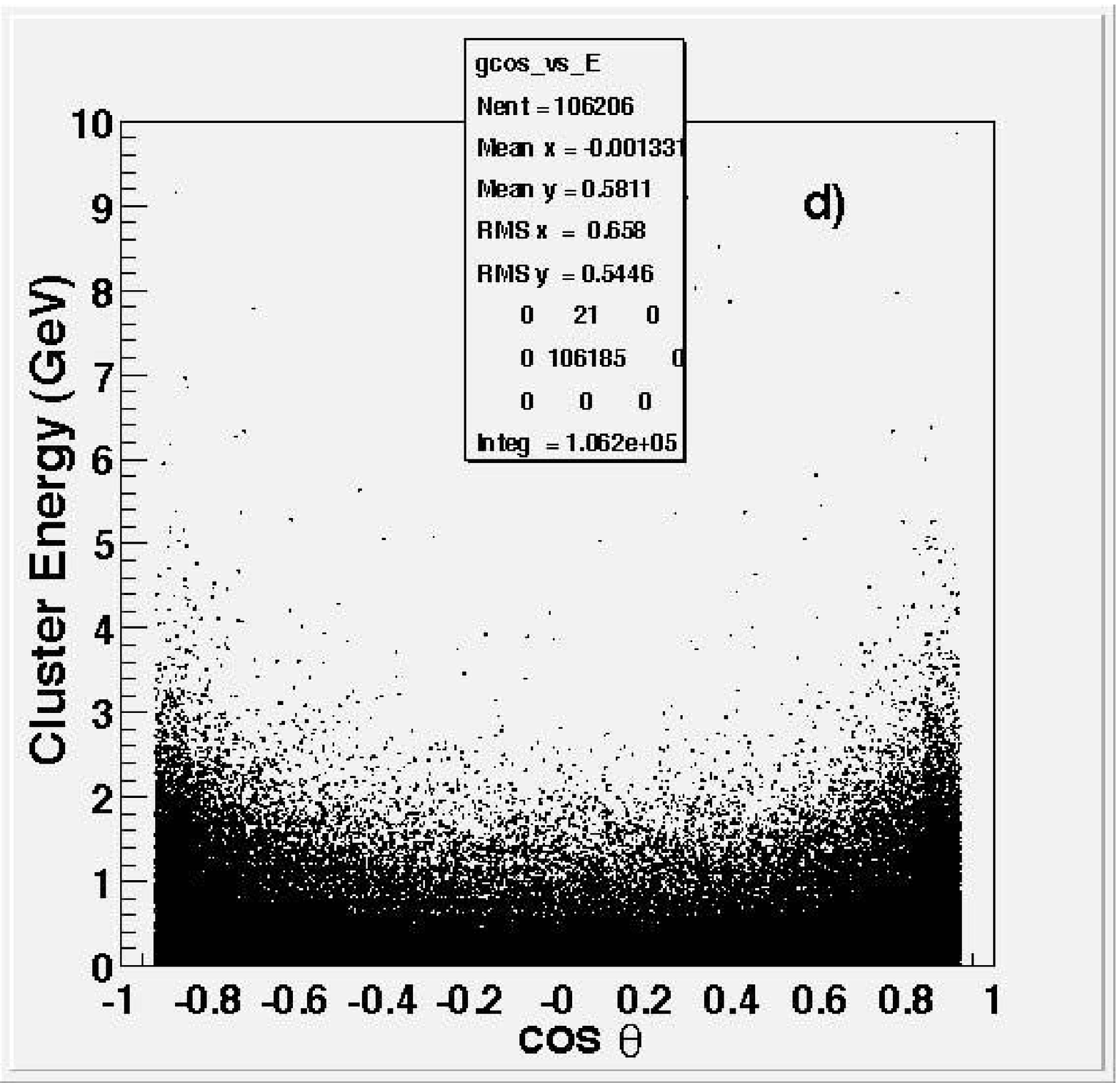}
\includegraphics{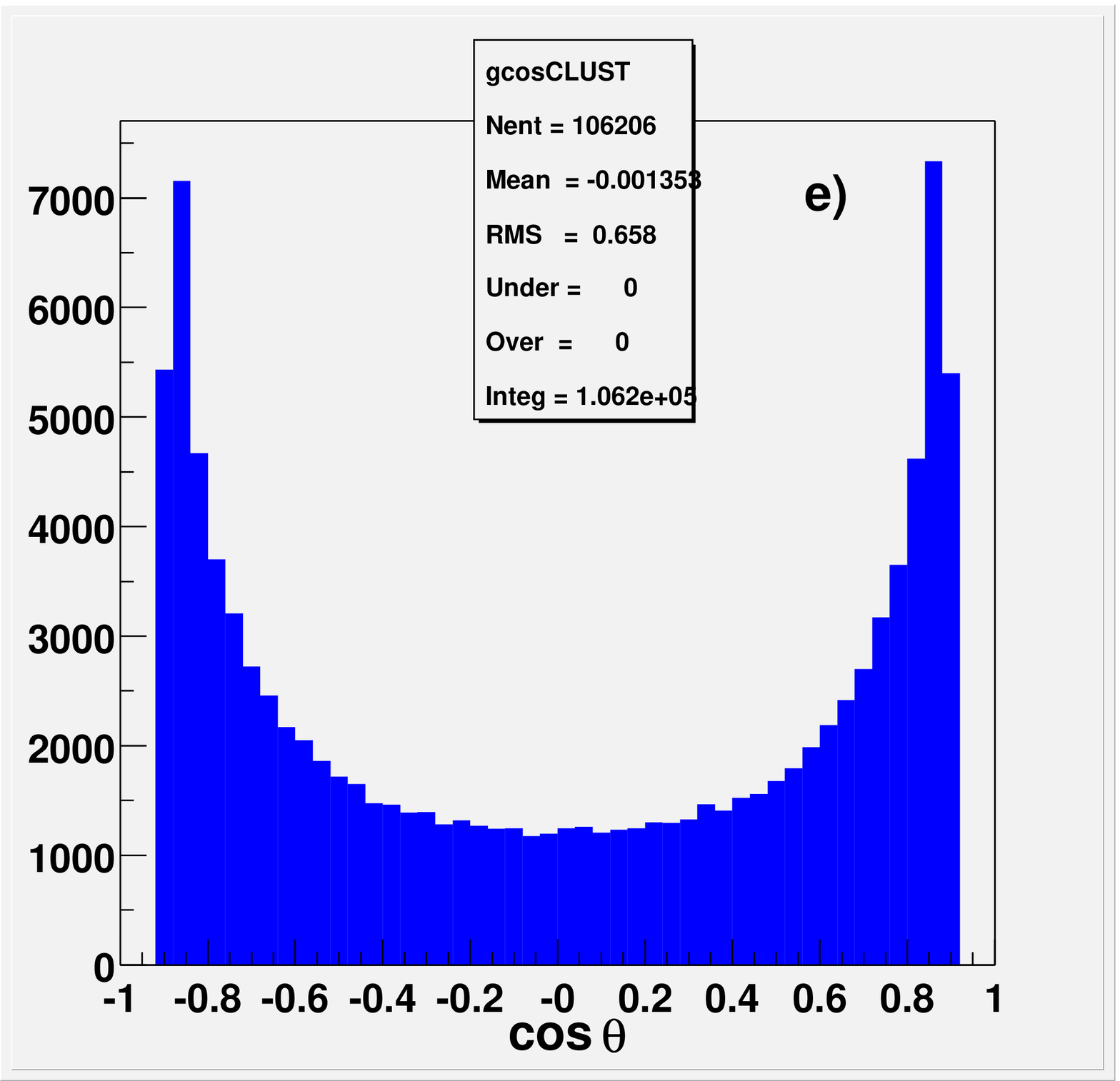}
\includegraphics{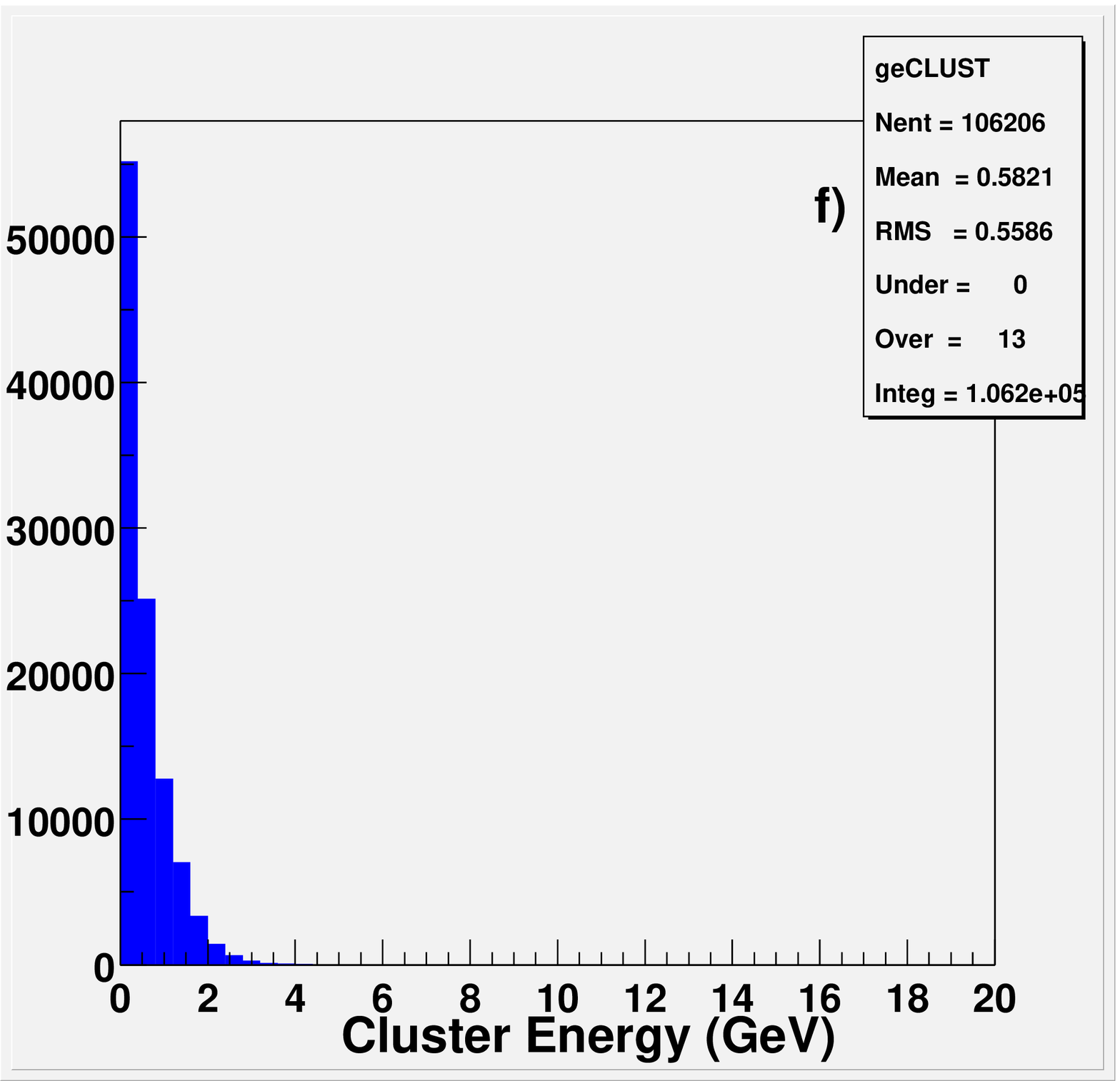}}
\caption{\label{fig:500resolve}
Tracks and Shower contributing to the resolved photon background
for $\rtsee=500$~GeV
a) Momentum versus $\cos\theta$ distribution for tracks with $p>0.2$~GeV/$c$.
b) and c) are the horizontal and vertical projections of a).
d) Energy versus $\cos\theta$ distribution for showers with $E>0.1$~GeV.
e) and f) are the horizontal and vertical projections of d).
}
\end{center}\end{figure}

We consider the beam parameters for $\rtsee=150$~GeV and $\rtsee=500$~GeV in
Table~\ref{tab:beampar} and use the luminosity distributions 
shown in Fig.~\ref{fig:cainlums}.
We process the events through the 
LC Fast MC detector simulation within ROOT \cite{Brun:1997pa},
which includes calorimeter smearing and detector configuration as
described in Section 4.1 of Chapter 15 of Ref.~\cite{Abe:2001nr}.
Our preliminary studies find that the two-photon cross section
is dominated by the twice resolved process at both 150~GeV and 500~GeV.
We find the once resolve event yield to be approximately 
400 and 100 times larger
than the direct cross section for 150~GeV and 500~GeV, respectively,
and we find the twice resolve event yield to be approximately
1500 and 500 times larger
than the direct cross section for 150~GeV and 500~GeV, respectively.
However, these cross sections have large uncertainties which we discuss below.
Most of the products of the $\gamma\gamma \to$~hadrons will be produced
at small angles relative to the photon beam and will escape down the
beam pipe undetected. We are interested in the decay products that
enter the detector and we consider
only tracks and showers with $|\cos\theta|<0.9$ in the laboratory
frame and we require 
tracks have to have momentum greater than 200~MeV and showers
must have energy greater than 100~MeV.
The resulting track and shower energy distributions integrated over 
20,000 beam crossings for $\rtsee=150$~GeV and 1,000 beam crossings for $\rtsee=500$~GeV are shown in
Fig.~\ref{fig:500resolve}.
Experimental, theoretical and modeling errors have not yet been evaluated for
these distributions.
\par
Future studies of the physics possibilities of a $\gaga$~collider should
include the impact of resolved photons on the event reconstruction. 
The appropriate number of beam crossing to integrate over has not been
determined. It is generally assumed that the $\gamma\gamma$ and $e^+e^-$
detectors will be the same or at least have comparable performance.
At $e^+e^-$, the plan is to integrate over 100's to 1000's of beam crossings.
This depends, of course, on the choice of detector technology as well as 
the bunch structure of the electron beam. At a $\gamma\gamma$~collider 
experiment, the desire to minimize the resolved photon backgrounds 
may drive the detector design to
integrate over as few beam crossings as possible. Thus the assumption
that the $e^+e^-$ and $\gamma\gamma$ detectors will be based on the same
technology or have comparable performance may not be valid. An important
distinction between the TESLA and NLC machine designs 
is the time between bunch crossings, 337~ns and 2.8~ns\footnote{The NLC-$e^+e^-$design has 1.4~ns bunch spacing -- see the caption of Table~\ref{tab:beampar}
for discussion.}
, respectively.
The authors plan to study the (hopefully) incremental difference between
the design and performance of a detector for the $\gam\gam$ and $e^+e^-$
interaction regions. 

Large uncertainties plague the estimated once and twice resolved photon
cross sections. Ref.~\cite{butterworthpaper} provides 
an excellent discussion of the theoretical
and experimental challenges of determining the photon structure.
Most recent experimental data is from HERA~\cite{hera} and LEP-II~\cite{lep}.
Large uncertaintes exist for small $x_\gamma$ and significant uncertainties
exist for large $x_\gamma$.
The QCD stucture function is probed by examining high $E_t$ jets. Extrapolating
to smaller $E_t$ introduces additional uncertainty.
The situation may be improved by data obtained from a proposed 
$\gamma\gamma$ engineering run at SLC~\cite{ourproposal}.

The impact of resolved photon backgrounds on the physics reach of a 
$\gamma\gamma$ collider has yet to be determined. Although unevaluated (and
likely large) uncertainties in the expected track and shower momentum/energy
distribution and normalization remain, we have incorporated the resolved photon
background levels presented here into our analysis of $\gam\gam \to h \to \gam\gam$. The preliminary results are promising - see Section~\ref{sec:hgg}.

\begin{figure}\begin{center}
\resizebox{\textwidth}{!}{
\includegraphics{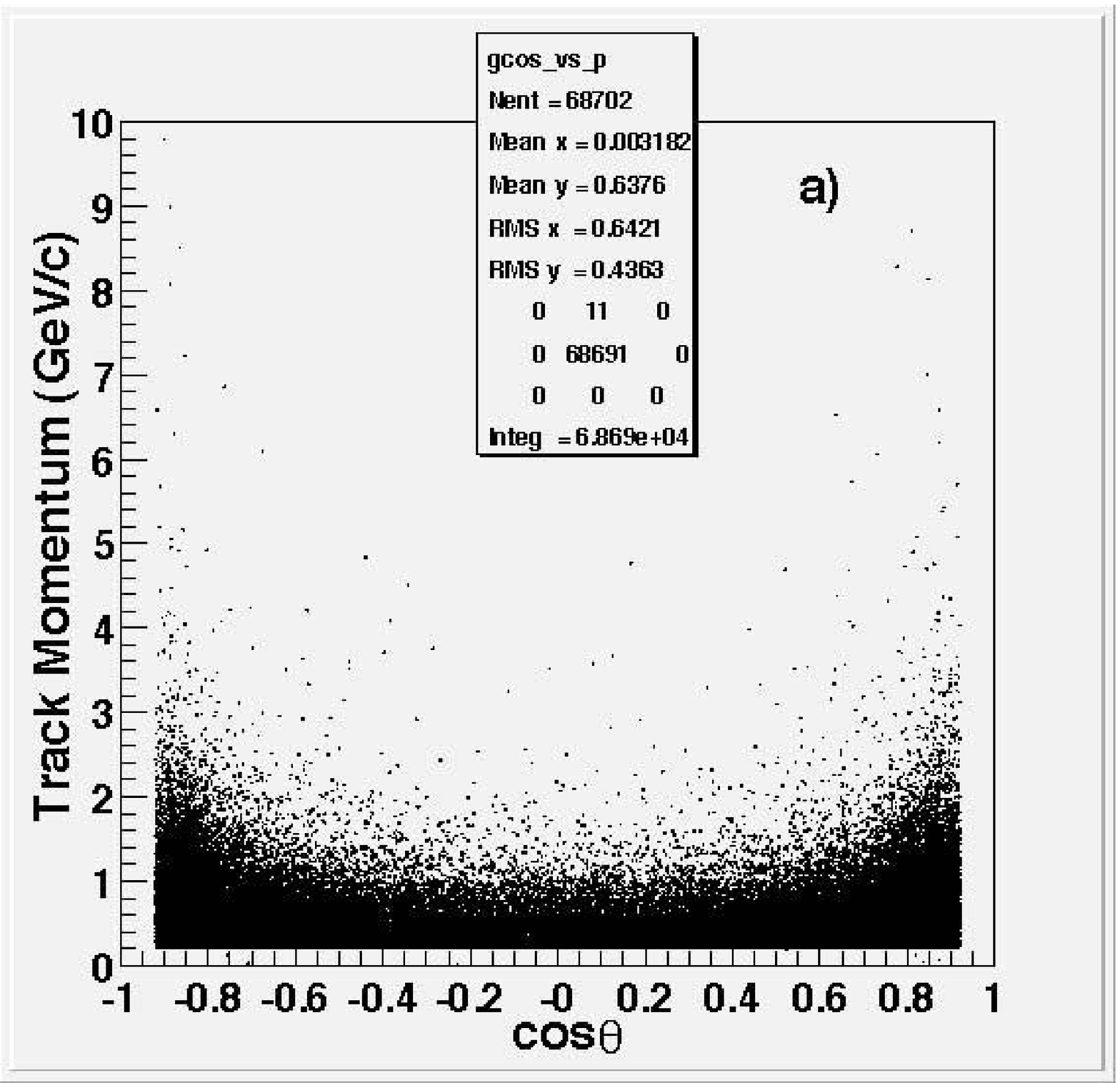}
\includegraphics{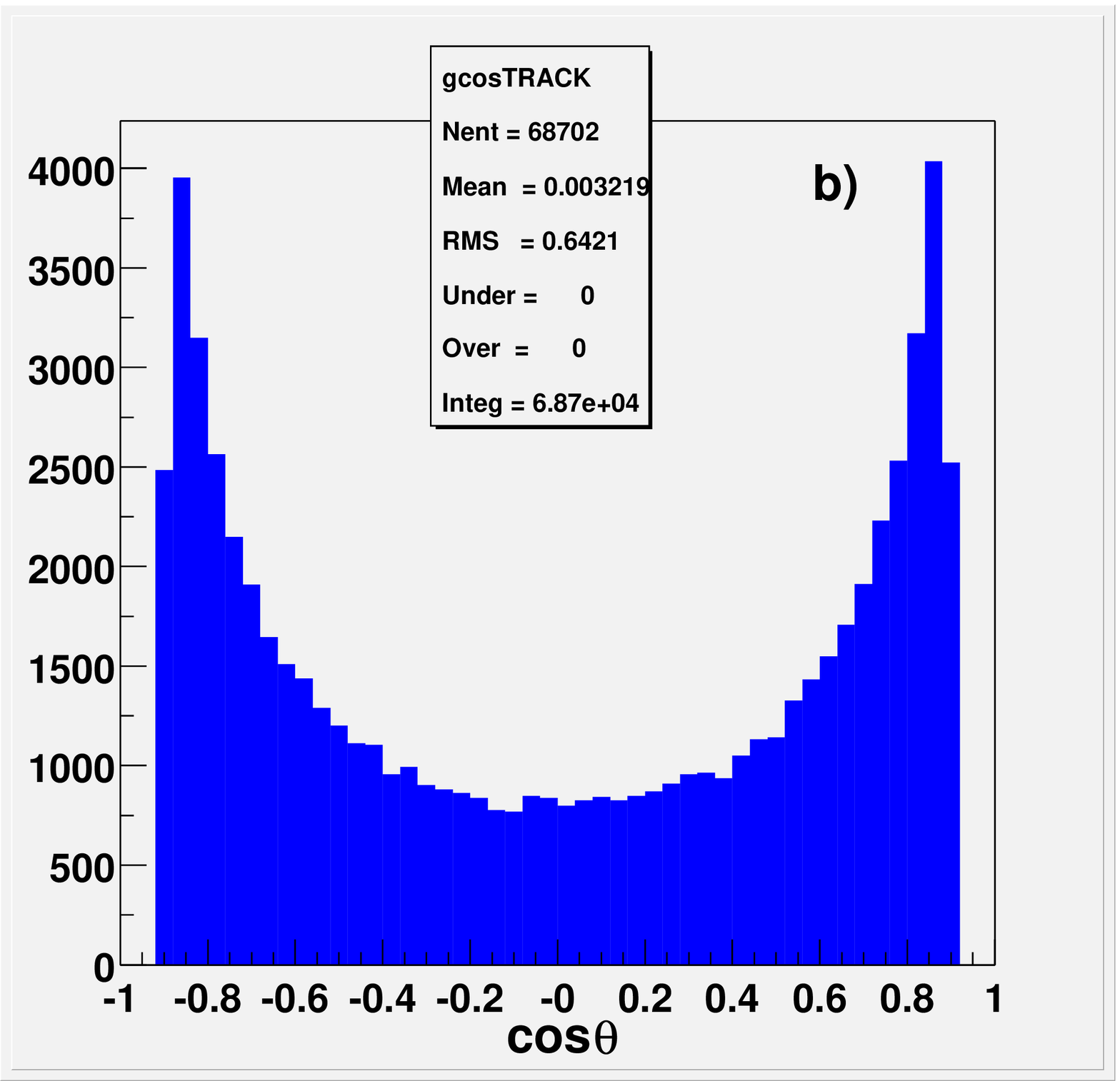}
\includegraphics{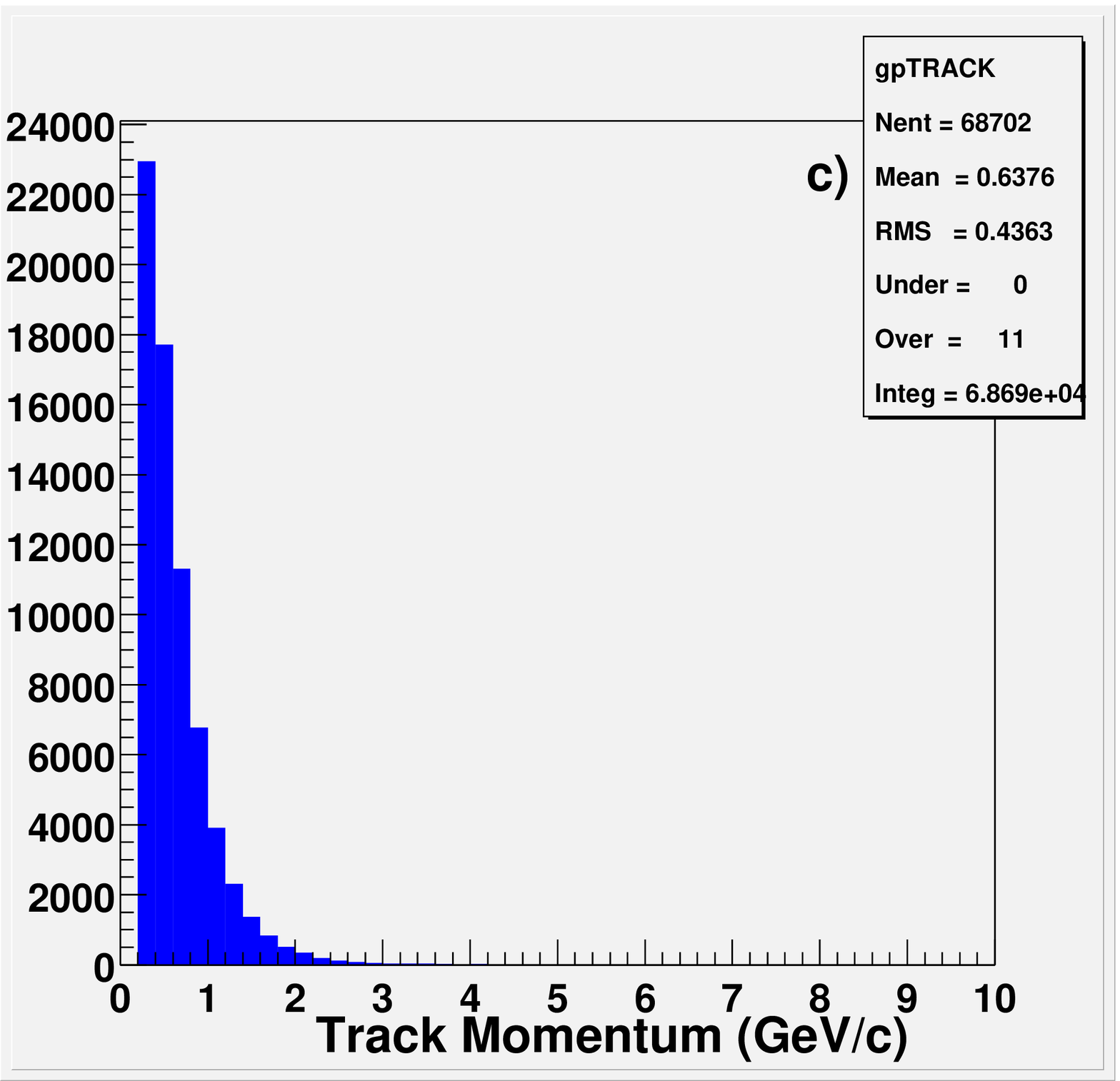}}
\resizebox{\textwidth}{!}{
\includegraphics{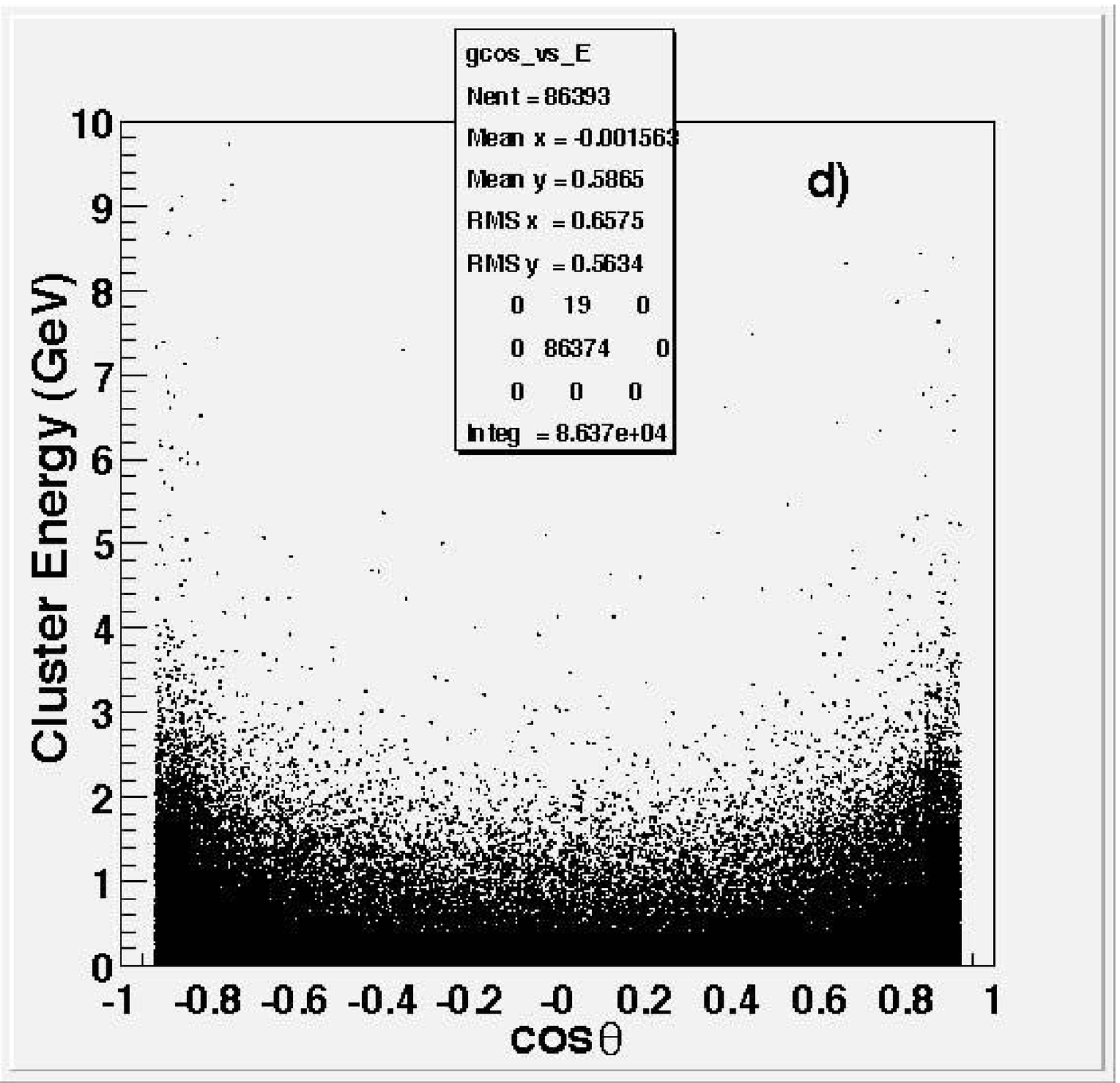}
\includegraphics{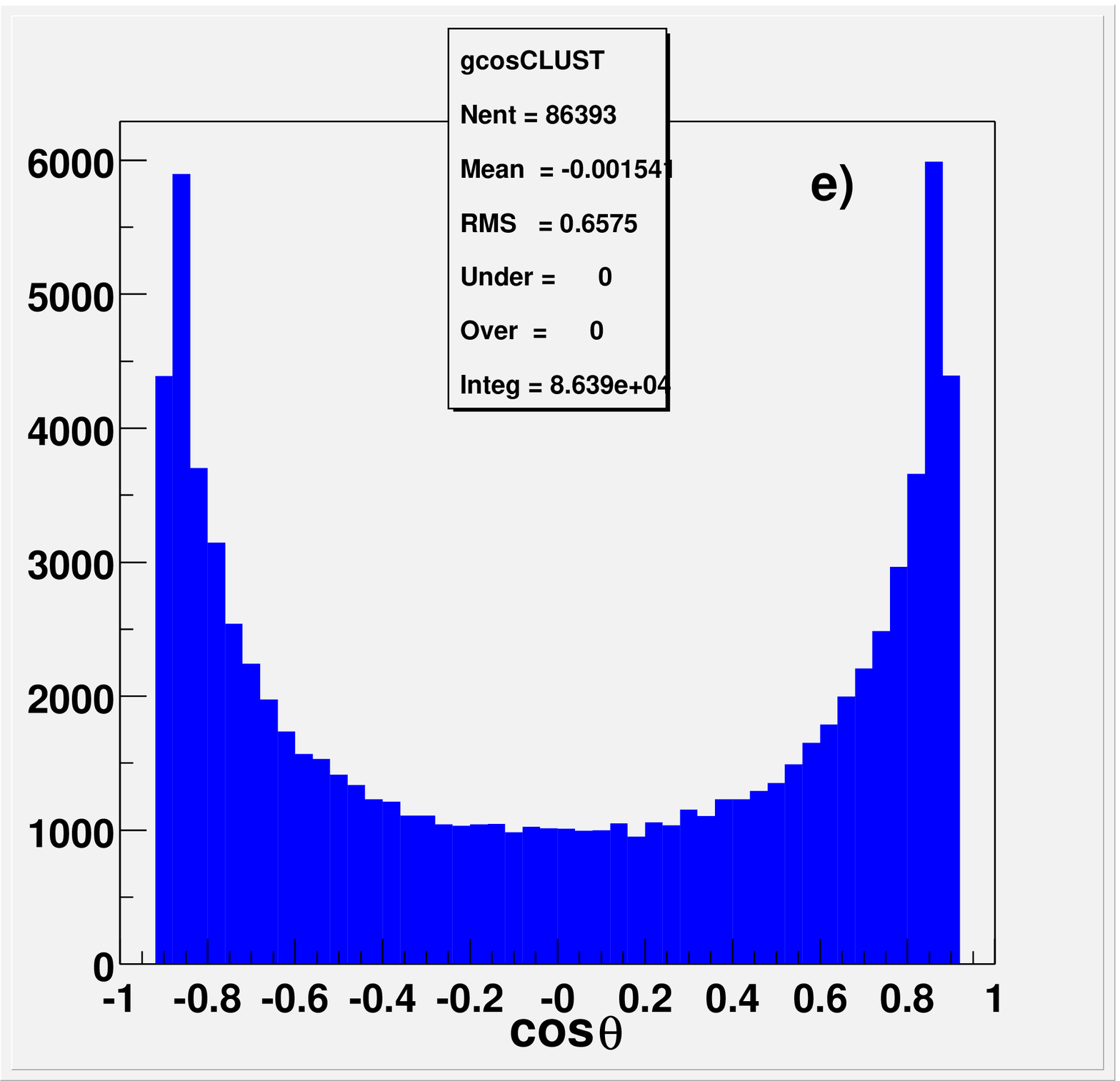}
\includegraphics{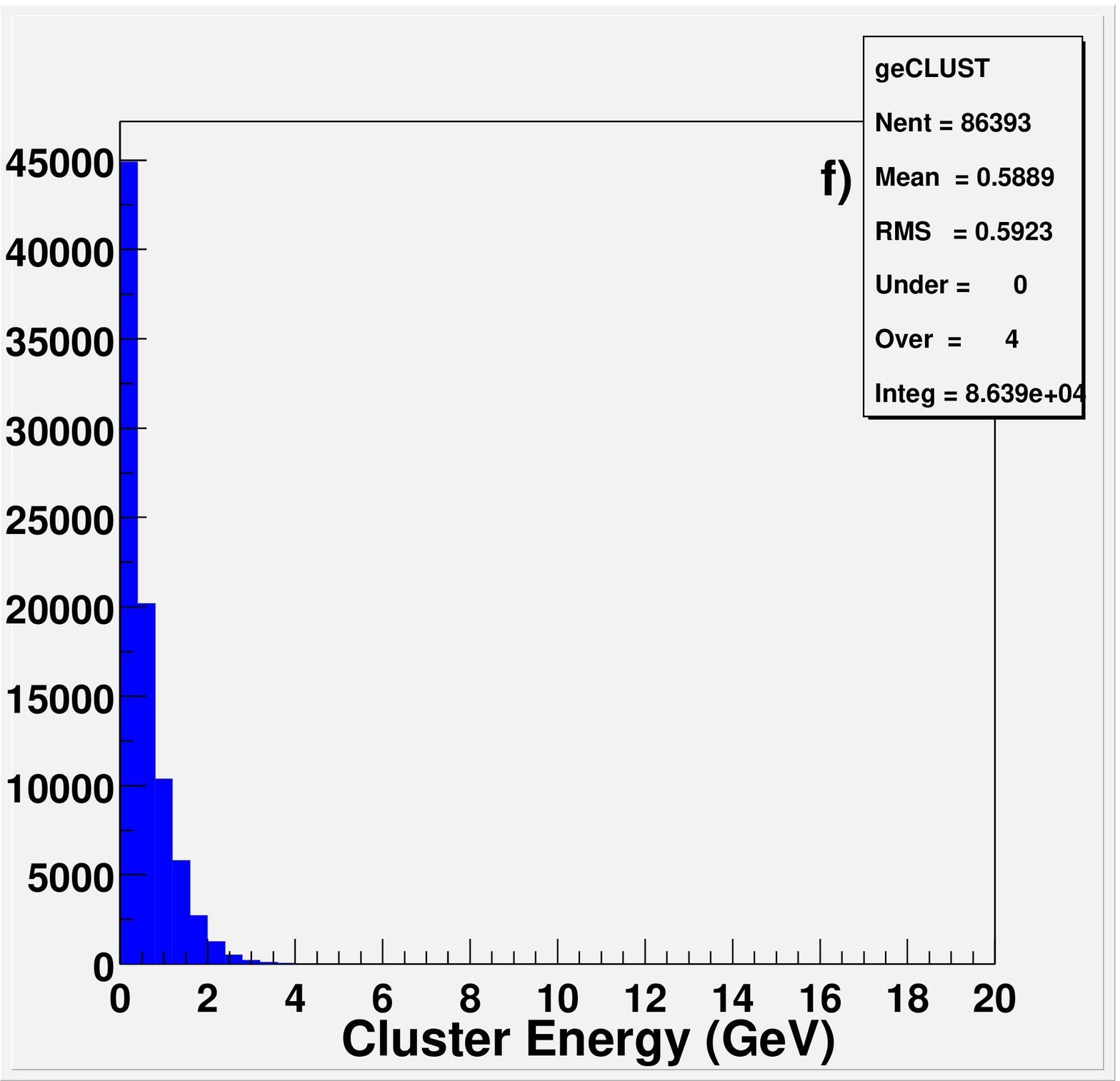}}
\caption{\label{fig:150resolve}
Tracks and Shower contributing to the resolved photon background
for $\rtsee=150$~GeV
a) Momentum versus $\cos\theta$ distribution for tracks with $p>0.2$~GeV/$c$.
b) and c) are the horizontal and vertical projections of a).
d) Energy versus $\cos\theta$ distribution for showers with $E>0.1$~GeV.
e) and f) are the horizontal and vertical projections of d).
}
\end{center}\end{figure}

%The physics, backgrounds and radiation environment are not identical
%in the $\gamma\gamma$ and $e^+e^-$ interaction regions; however,
%when studying the physics capabilities of the two-photon interaction
%region, it is common to assume the detector design and performance 
%of $e^+e^-$ detectors.  We discuss the required modifications of the
%nominal $e^+e^-$ detector design to accomodate the backgrounds and 
%radiation of the two-photon interaction region.

%----------------------------------------------------------------
\section{\label{sec:summary}Conclusion}
\par
Several new results in the physics to be studied with $\gaga$~colliders
have been briefly described.  Most of these come from work in progress,
and will be refined and extended over the next several months.
\par
We believe that the next most important step is the study and understanding
of resolved photon backgrounds, which appear to be formidable and possibly quite pernicious.
They may force the detector design to differ significantly from the
one currently under development for the $\epem$ linear collider.
To this end, a proposal to perform a proof-of-principle experiment
at SLAC is under way, and a workshop on the subject is planned for
November,~2002.
\par
In addition to understanding machine and background issues, it is time
to take  a closer look at the interplay of results expected from a
high energy $\epem$~collider and a $\gaga$~collider.  Discussions are
already planned for LC2002, and should be continued at other workshops
over the next year.  A goal of such discussion should be a straw proposal
for a combined run plan for, a 500~GeV machine with both $\epem$ and
$\gaga$~interaction regions active.

%---------------------------

\acknowledgments
We are grateful to Jon Butterworth, Marcela Carena, Jeff Gronberg, Tony Hill,
Steve Mrenna, and Michael Peskin for helpful discussions.
Fermilab is operated by Universities Research Association Inc.\
under contract no.~DE-AC02-76CH03000 with the U.S. Department of
Energy.
A portion of this work was performed under the auspices of the 
U.S. Department of Energy 
by the University of California, Lawrence Livermore National
Laboratory under Contract No.W-7405-Eng.48.
Other portions were supported by the Illinois Consortium for
Accelerator Research, agreement number~228-1001.
B.G. is supported in part by the State Committee for Scientific
Research under grant 5~P03B~121~20 (Poland). J.F.G. is supported
by the U.S. Department of Energy and by the Davis Institute for High
Energy Physics.

%-------------------------------------------------------


\begin{thebibliography}{99}
\bibitem{ourproposal}  We plan to demonstrate the technical feasibility of the
NLC $\gamma\gamma$ interaction region design.
We propose to integrate a 1/2 scale NLC $\gamma\gamma$ interaction region 
into the SLC and generate two-photon luminosity by Compton scattering  30~GeV
beams from the SLC off a 0.1 J short pulse laser.
\end{thebibliography}
\end{document}